\documentclass{elsart}
\usepackage{amssymb}

\usepackage{graphicx}

\begin{document}

\begin{frontmatter}



\title{Magnetic nanoparticles as many-spin systems}


\author[kac]{H. Kachkachi}
\author[gar]{and D. A. Garanin}
\address[kac]{Laboratoire de Magn\'{e}tisme et d'Optique,
Universit\'e de Versailles St. Quentin, \\
45 av. des Etats-Unis, 78035 Versailles, France}
\address[gar]{Institut f\"ur Physik, Johannes-Gutenberg-Universit\"at,
 D-55099 Mainz, Germany}

\begin{abstract}
We present a review of recent advances in the study of many-body
effects in magnetic nanoparticles. Considering classical spins on
a lattice coupled by the exchange interaction in the presence of
the bulk and surface anisotropy, we investigate the effects of
finite size, free boundaries, and surface anisotropy on the
average and local magnetization for zero and finite temperatures
and magnetic fields. Superparamagnetism of magnetic particles
necessitates introducing two different, induced and intrinsic,
magnetizations. We check the validity of the much used relation
between them within different theoretical models. We show that
the competition between the exchange and surface anisotropy leads
to spin canting dependent on the orientation of the average
magnetization with respect to the crystallographic axes and thus
to a second-order effective anisotropy of the particle.
We have also investigated the switching mechanism of the magnetization upon
varying the surface anisotropy constant.
Some cases of more realistic particles are also dealt with.
\end{abstract}

\end{frontmatter}


\section{Introduction}

\label{sec:introduction} Magnetic nanoparticles, or nanoscale magnetic
systems, have generated continuous interest since late 1940s as the
investigation of their properties turned out to be challenging from both
scientific and technological point of view. In 1949, in a pioneering work
\cite{nee49crnee49ang}, N\'{e}el set the pace towards understanding of the
magnetic behavior of nanoparticles, leading to an important development of
fundamental theories of magnetism and modeling of magnetic materials, as
well as remarkable technological advances, e.g., in the area of information
storage and data processing, fostering the development of magnetorecording
media with increasingly higher densities. Nanoparticles, as compared to bulk
materials, possess very important novel properties such as enhanced
remanence and giant coercivity, as well as exponentially slow relaxation at
low temperature due to anisotropy barriers, which ensures great stability of
the information stored. However, nanoparticles become \emph{superparamagnetic
} \cite{nee49crnee49ang} at finite temperatures for very small sizes, and
this is an impairment to the information storage. On the other hand,
discovery of superparamagnetism that results from thermally activated
crossing of the anisotropy energy barrier by the magnetic moment of the
particle opened a rich area to the application of nonequilibrium statistical
mechanics. There are mainly two types of nanoparticle samples: i) Assemblies
of, e.g., cobalt, nickel or maghemite, nanoparticles with volume
distribution and randomly oriented axes of magnetocrystalline anisotropy;
ii) Isolated single particles of cobalt or nickel that can be probed by the
micro-SQUID technique \cite{wer01acp}. While most of the experiments have been
done on assemblies, isolated particles are more important both as units of
information storage and as a physical system.

In the investigation of the static properties of magnetic nanoparticles a
great deal of work up to date has been based on the Monte Carlo (MC)
technique. In addition to numerous simulations of the Ising model, this
technique has been used with the more adequate classical Heisenberg model to
simulate idealized isotropic models with simple cubic (sc) lattice and
spherical shape in Ref.~\cite{wil74}. Magnetic nanoparticles with realistic
lattice structure were recently simulated in Ref.~\cite{kacetal00epjb}
taking into account the surface anisotropy (SA) and DDI.

On the other hand, in most of theoretical approaches to the dynamics of a
small magnetic particle the latter is considered as a single magnetic moment.
This is the \emph{one-spin approximation} that is only valid for particles that
are not too large and thus are single-domain, and not too small to be free
from surface effects.
Letting apart spin tunneling (see, e.g., Ref. \cite{chutej98book}) that becomes
important for extremely small sizes such as those of molecular magnets, the
magnetic moment can overcome the anisotropy-energy barrier and thus reverse its
direction, at least in two ways \footnote{
It has quite recently been shown \cite{shuetal03prl}, experimentally and
theoretically, that efficient magnetization switching can be triggered by
transverse field pulses of a duration that is half the precession period.}:
Either under applied magnetic field that suppresses the barrier, or via
thermal fluctuations. The former, at zero temperature, is well described for
particles with the uniaxial anisotropy by the Stoner-Wohlfarth model \cite{stowoh48}.
Thermally activated crossing of the energy barrier is described by the
N\'{e}el-Brown model \cite{bro63prbro79ieee} and its extensions (see, e.g.,
Refs. \cite{cofgarmcc01}). At elevated temperatures, rotation of the
magnetization in materials with strong anisotropy is always accompanied by
changing its magnitude.
This results in a shrinking of the Stoner-Wohlfarth
astroid as described by the modified Landau theory \cite{kacgar01physa}, and
(qualitatively) confirmed by experiments \cite{wer01acp}.

Both Stoner-Wohlfarth and N\'{e}el-Brown models have been confirmed by
experiments on individual cobalt particles \cite{wer01acp}. However for
magnetic particles with strong surface anisotropy magnetization switching
occurs as the result of successive switching of individual (or clusters of)
spins inside the particle \cite{kacdim02prbjap}. Such deviations from the
one-spin approximation have been observed in metallic particles \cite
{chesorklahad95prb}, \cite{resetal98prb}, and ferrite particles
\cite{kodber99prb}, \cite{ricetal91jap}.
Deviations from the one-spin
approximation and temperature effects lead to the absence of magnetization
saturation at high fields \cite{chesorklahad95prb}, \cite{ezzirthesis98},
\cite{resetal98prb}, shifted hysteresis loops after cooling in field, and
field dependence of the magnetization at very low temperatures. The latter
effect has been clearly identified in dilute assemblies of maghemite
particles \cite{troetal00jmmm} of $4$ nm in diameter. In addition, aging
effects have been observed in single particles of cobalt and have been
attributed to the oxidation of the sample surface into antiferromagnetic CoO
(see \cite{wer01acp} and references therein.). It was argued that the
magnetization reversal of a ferromagnetic particle with antiferromagnetic
shell is governed by two mechanisms that are supposed to result from the
spin frustration at the core-shell interface of the particle. Some of the
above-mentioned novel features are most likely due to magnetic disorder at
the surface which induces a canting of spins inside the particle, or in
other words, an inhomogeneous magnetic state. This effect was first observed
with the help of M\"{o}ssbauer spectroscopy by Coey \cite{coeprl71} and
later by Morrish and Haneda (see \cite{hancap87} for a review), and later
by Pren\'e et al. \cite{prehi94} (see also the recent article \cite{morup}).
To sum up, the picture of a
single-domain magnetic particle with all spins pointing into the same
direction is no longer valid when one considers the effect of misaligned
spins on the surface, which makes up to $50\%$ of the total volume in a
particle of $4$ nm in diameter.

One of our goals is thus is to understand the effect of surfaces on the
thermodynamics and magnetization profiles in small systems, and subsequently
upon their dynamics. This requires a microscopic approach to account for the
local environment inside the particle, microscopic interactions such as
spin-spin exchange, DDI, and the magneto-crystalline bulk and surface
anisotropy. As this task is difficult owing to the large number of degrees of
freedom involved, one has to gain a sufficient understanding of static
properties before proceeding to the dynamics. There are three main effects that
distinguish magnetic particles from bulk magnets and that were investigated
in a series of our recent publications:

\begin{enumerate}
\item  Finite-size effect in isotropic magnetic particles with idealized
periodic boundary conditions. The spin-wave spectrum of such particles is
discrete and there is the mode with $\mathbf{k}=0$ that corresponds to the
global rotation of the particle. As a result, the standard spin-wave theory
fails and one has to distinguish between $\emph{induced}$ and \emph{intrinsic}
magnetizations.

\item  Boundary effect, i.e., pure effect of the free boundary conditions at
the surface in the absence of the surface anisotropy. This effect leads to
the decrease of the particle's intrinsic magnetization and it makes the
latter inhomogeneous at $T\neq 0$.

\item  Effect of surface anisotropy that changes the ground state of the
particle and makes the intrinsic magnetization inhomogeneous even at $T=0$.
\end{enumerate}

This paper is organized as follows: In Sec. \ref{sec:basic} we describe the
Hamiltonian and introduce the basic notions of the induced and the intrinsic
magnetizations. In Sec.~\ref{sec:mspwtse}, we first study inhomogeneities in
small magnetic particles of a box shape induced by pure boundary effects in
the absence of surface anisotropy, i.e., the effect of free boundary
conditions (fbc), and compare their influence with that of the finite-size
effects in an idealized model with periodic boundary conditions (pbc). We
consider spins as $D$-component classical vectors. At first we present
analytical and numerical results in the whole range of temperatures in the
limit $D\rightarrow \infty $ where the problem simplifies. Then for the
classical Heisenberg model, $D=3,$ at low temperatures we formulate the
modified spin-wave theory accounting for the global rotation of the
particle's magnetization. The results are compared with those of our MC
simulations. In Sec.~\ref{sec:mspwse} we consider round-shaped systems and
include surface anisotropy. In the case of the SA much weaker than the
exchange interaction we study the problem perturbatively in small deviations
from the perfectly ordered, collinear state. Then we investigate the
hysteretic properties and the behavior of the magnetization as a function
of temperature and applied field by MC simulations. The last section
summarizes the results and points out open problems.

\section{\label{sec:basic}Basic relations}

\subsection{The Hamiltonian}

\label{sec:Energy}

Within the classical approximation it is convenient to represent the atomic
spin as the three-component spin vector $\mathbf{s}_{i}$ of unit length on
the lattice site $i$. We will consider the Hamiltonian that in general
includes the exchange interaction, magneto-crystalline anisotropic energy,
Zeeman energy, and the energy of dipolar interactions (DDI)
\begin{equation}
\mathcal{H}=-\frac{1}{2}\sum\limits_{ij}J_{ij}\mathbf{s}_{i}\cdot \mathbf{s}
_{j}-\mu _{0}\mathbf{H}\cdot \sum\limits_{i}\mathbf{s}_{i}+\mathcal{H}_{%
\mathrm{an}}+\mathcal{H}_{\mathrm{DDI}},  \label{hamgeneral}
\end{equation}
where $\mu _{0}=g\mu _{B}S$ and $S$ is the value of the atomic spin.
For materials with uniaxial anisotropy $\mathcal{H}_{\mathrm{an}}$ in
Eq.~(\ref{hamgeneral}) reads
\begin{equation}
\mathcal{H}_{\mathrm{an}}^{(\mathrm{uni})}=-\sum\limits_{i}K_{i}(\mathbf{s}
_{i}.\mathbf{e}_{i})^{2},  \label{uaa}
\end{equation}
with easy axis $\mathbf{e}_{i}$ and constant $K_{i}>0$. This anisotropy
model can be used to describe the surface effect if one attributes the same
easy axis and the same anisotropy constant $K_{c}$ for all core spins and
different easy axes and anisotropy constants to surface spins. Within the
simplest transverse surface anisotropy (TSA) model all surface spins have
the same anisotropy constant $K_{s},$ whereas their easy axes are
perpendicular to the surface, see. e.g., Refs. \cite{aha96,bro63mic,shiphd99}
. More realistic is N\'{e}el's surface anisotropy (NSA) model \cite{nee54jpr}
,
\begin{equation}
\mathcal{H}_{\mathrm{an}}^{(\mathrm{NSA)}}=-L\sum\limits_{i}\sum%
\limits_{j=1}^{z_{i}}(\mathbf{s}_{i}\cdot \mathbf{e}_{ij})^{2},\qquad
\mathbf{e}_{ij}\equiv \mathbf{r}_{ij}/r_{ij},\qquad \mathbf{r}_{ij}\equiv
\mathbf{r}_{i}-\mathbf{r}_{j},  \label{NSA}
\end{equation}
where $z_{i}$ is the coordination number of site $i$ that for the surface
atoms is smaller than the bulk value $z$ and $\mathbf{e}_{ij}$ is the unit
vector connecting the site $i$ to its nearest neighbors $j.$ One can check
that for the simple cubic (sc) lattice the contributions from the bulk spins
in (\ref{NSA}) $\sim \mathbf{s}_{i}^{2}=1$ are irrelevant constants.

For materials with magneto-crystalline cubic anisotropy $\mathcal{H}_{%
\mathrm{an}}$ reads
\begin{equation}
\mathcal{H}_{\mathrm{an}}^{(4)}=-K^{(4)}\sum\limits_{i}\left(
s_{ix}^{4}+s_{iy}^{4}+s_{iz}^{4}\right) .  \label{ca_xyz}
\end{equation}
For $K^{(4)}>0$, the energy $\mathcal{H}_{\mathrm{an}}^{(4)}$ has minima for
six orientations of the type [100] and maxima for eight orientations of type
[111].

If one discards $\mathcal{H}_{\mathrm{an}}^{(4)}$ and $\mathcal{H}_{\mathrm{%
DDI}},$ the Hamiltonian (\ref{hamgeneral}) can be generalized for $D$%
-component spin vectors. This is useful as in the limit $D\rightarrow \infty
$ the problem simplifies while retaining important physics, see below.

\subsection{\label{sec:Magnetization}Magnetization of finite systems}

Magnetic particles of \emph{finite size} do not show magnetic ordering at
nonzero temperatures at $H=0$ as the global magnetization of the particle
can assume all possible directions (superparamagnetism). It is thus
convenient to define two magnetizations, $m$ and $M,$ the first being the
magnetization induced by the magnetic field and the second being a measure
of the short-range order in the particle. Omitting the factor $\mu _{0}$
that can be restored later, we first define the magnetization of a
microscopic spin configuration
\begin{equation}
\mathbf{M}=\frac{1}{\mathcal{N}}\sum_{i}\mathbf{s}_{i},  \label{MvecDef}
\end{equation}
where $\mathcal{N}$ is the number of magnetic atoms in the system. The
thermodynamic average of $\mathbf{M}$ yields what we call the \emph{induced
magnetization}
\begin{equation}
\mathbf{m}=\langle \mathbf{M}\rangle =\frac{1}{\mathcal{N}}\sum_{i}\langle
\mathbf{s}_{i}\rangle .  \label{mdef}
\end{equation}
The \emph{intrinsic magnetization} is related to the spin correlation
function:
\begin{equation}
M=\sqrt{\left\langle \mathbf{M}^{2}\right\rangle }=\sqrt{\left\langle \left(
\frac{1}{\mathcal{N}}\sum\limits_{i}\mathbf{s}_{i}\right) ^{2}\right\rangle }%
=\frac{1}{\mathcal{N}}\sqrt{\sum\limits_{ij}\left\langle \mathbf{s}_{i}\cdot
\mathbf{s}_{j}\right\rangle }.  \label{Mdef}
\end{equation}

If the temperature is low and there is no surface anisotropy, all spins in
the particle are bound together by the exchange interaction and $\mathbf{M}$
behaves as a rigid ``giant spin'', $|\mathbf{M|}\cong M\mathbf{\cong }1$. If
a magnetic field $\mathbf{H}$ is applied, $\mathbf{M}$ exhibits an average
in the direction of $\mathbf{H,}$ which leads to a nonzero value of the
induced magnetization $\mathbf{m.}$ For isotropic $D$-component vector
models (including the Ising model, $D=1)$ \cite{sta68prlpr} the latter is
given by the well known formula

\begin{equation}
m=MB_{D}(Mx),\qquad x\equiv \mathcal{N}H/T,  \label{spmrelation}
\end{equation}
where $B_{D}(x)$ is the Langevin function [$B_{3}(x)=\coth x-1/x$ for the
isotropic Heisenberg model and $B_{1}(x)=\tanh x$ for the Ising model]. An
important question is whether Eq. (\ref{spmrelation}) remains valid at
elevated temperatures where $M=M(T,H)$. We have shown that the \emph{%
superparamagnetic relation,} Eq. (\ref{spmrelation}), becomes exact for $%
D\rightarrow \infty $ but otherwise it contradicts the exact relation
\begin{equation}
M^{2}=m^{2}+\frac{dm}{dx}+\frac{(D-1)m}{x}.  \label{Mviam}
\end{equation}

One also can introduce the local intrinsic magnetization $M_{i}$ according
to
\begin{equation}
M_{i}=\frac{1}{M}\left\langle \mathbf{s}_{i}\cdot \frac{1}{\mathcal{N}}%
\sum\limits_{j}\mathbf{s}_{j}\right\rangle ,\qquad \frac{1}{\mathcal{N}}%
\sum_{i}M_{i}=M.  \label{MLocalDef}
\end{equation}
This quantity is smaller near the boundaries of the particle than in the
core for the model with fbc because of boundary effects at $T>0.$

\section{\label{sec:mspwtse}Nanoparticle as a multi-spin system: finite-size
vs boundary effects}

Finite-size magnetic systems with free boundary conditions (fbc) present a
spatially inhomogeneous many-body problem. In this section we shall only
deal with finite-size versus boundary effects, leaving the more profound
effects of the surface anisotropy for the next section. One of the
interesting problems here is the interplay between boundary effects due to fbc
and the ``pure'' finite-size effects. In systems of hypercubic shape, the
latter can be singled out by using artificial periodic boundary conditions
(pbc). The standard mean-field approximation (MFA) and spin-wave theory
(SWT) are inappropriate for finite systems because of the Goldstone mode
associated with the global rotation of the magnetic moment in zero field.
Appropriate improvements include the so-called $D\rightarrow \infty $ model
(see Sec. \ref{sec:ASM}) operating at all temperatures and the modified
spin-wave theory for finite magnets at low temperatures (see Sec.~\ref
{sec:SWTMC}). Also the MC routine should incorporate global rotations of
spins, in addition to the Metropolis algorithm of individual spin rotations.

\subsection{\label{sec:ASM}$D\rightarrow \infty $ model}

\subsubsection{The model}

One can improve upon the MFA by taking into account correlations in a wide
temperature range for bulk and finite magnets by replacing 3-component spin
vectors in (\ref{hamgeneral}) by $D$-component ones and taking the limit $%
D\rightarrow \infty $. This model was introduced by Stanley \cite{sta68prlpr}
who showed that in the bulk its partition function coincides with that of
the exactly solvable spherical model (SM) \cite{berkac52}. On the other
hand, for spatially inhomogeneous and anisotropic systems the $D\rightarrow
\infty $ \ model is the only physically acceptable model of both (see, e.g.,
Refs. \cite{gar97zpb96jpa}). So far, the $D\rightarrow \infty $ \ model was
only applied to spatially inhomogeneous systems in the plane geometry \cite
{gar97zpb96jpa}. In Ref. \cite{kacgar01physa} we extended it to \emph{finite}
box-shaped magnetic systems with free and periodic boundary conditions. In
the MFA the Curie temperature of the $D$-component model is $T_{c}^{\mathrm{%
MFA}}=$ $J_{0}/D$, where $J_{0}$ is the zero Fourier component of $J_{ij}$.
It is convenient to use $T_{c}^{\mathrm{MFA}}$ as the energy scale and
introduce the dimensionless variables
\begin{equation}
\theta \equiv T/T_{c}^{\mathrm{MFA}},\qquad \mathbf{h\equiv H/}J_{0},\qquad
\lambda _{ij}\equiv J_{ij}/J_{0}.  \label{DefPar}
\end{equation}
For the nearest-neighbor (nn) interaction $J_{ij}$ with $z$ neighbors, $%
\lambda _{ij}$ is equal to $1/z$ if sites $i$ and $j$ are nearest neighbors
and zero otherwise. In the bulk the $D\rightarrow \infty $ model is
described by two coupled nonlinear equations for the magnetization $m$ and
the so called gap parameter $G$:
\begin{equation}
m=\frac{hG}{1-G},\qquad m^{2}+\theta GP(G)=1,\qquad P(G)=\int \!\!\!\frac{%
d^{3}\mathbf{k}}{(2\pi )^{3}}\frac{1}{1-G\lambda _{\mathbf{k}}},
\label{bulkEqs}
\end{equation}
where $P(G)$ is the lattice Green function and $\lambda _{\mathbf{k}}=\left(
\cos k_{x}+\cos k_{y}+\cos k_{z}\right) /3$ is the Fourier transform of $%
\lambda _{ij}.$ The Curie temperature is defined by $G=1$ and is $%
T_{c}=T_{c}^{\mathrm{MFA}}/W,$ i.e., $\theta _{c}=1/W,$ where $W\equiv P(0)$
is the Watson integral ($W=1.51639$ for the sc lattice). The system of
equations describing the inhomogeneous $D\rightarrow \infty $ model can be
obtained using the diagram technique for classical spin systems \cite
{garlut84d,gar94jsp96prb} in the limit $D\rightarrow \infty $ and
generalising the results of Ref.~\cite{gar97zpb96jpa} for spatially
inhomogeneous systems to include the magnetic field $\mathbf{h}=h\mathbf{e}%
_{z}$. This is a system of equations for the average magnetization $%
m_{i}\equiv \left\langle s_{zi}\right\rangle $ directed along the field$,$
gap parameter $G_{i},$ and correlation functions for the remaining
``transverse'' spin components labeled by $\alpha \geq 2$, i.e., $%
s_{ij}\equiv D\left\langle s_{\alpha i}s_{\alpha j}\right\rangle $ (all
transverse correlation functions are the same). This system of equations has the
form \begin{equation}
\sum_{j}\mathcal{D}_{ij}m_{j}=h,\qquad \sum_{j}\mathcal{D}_{ij}s_{jl}=\theta
\delta _{il},\qquad s_{ii}+m_{i}^{2}=1,  \label{DefMatr}
\end{equation}
where $\mathcal{D}_{ij}\equiv G_{i}^{-1}\delta _{ij}-\lambda _{ij}$ is the
Dyson matrix and $\delta _{il}$ is the Kronecker symbol. Solving this system
of equations consists in determining $m_{i}$ and $s_{ij}$ as functions of $%
G_{i}$ from the first two \emph{linear }equations and inserting the
solutions into the third nonlinear equation (the constraint equation) that
leads to a system of nonlinear equations for all $G_{i}$ that is in general
subject to numerical solution. Combining Eqs. (\ref{DefMatr}) results in $%
m^{2}+\theta m/(\mathcal{N}h)-M^{2}=0$ that yields
\begin{equation}
m=M\frac{2\mathcal{N}Mh/\theta }{1+\sqrt{1+(2\mathcal{N}Mh/\theta )^{2}}}%
=MB_{\infty }(\mathcal{N}MH/T),  \label{mvsM}
\end{equation}
where
\begin{equation}
M=\sqrt{\mathbf{m}^{2}+\frac{1}{\mathcal{N}^{2}}\sum\limits_{ij}s_{ij}}%
,\qquad B_{\infty }(\xi )=\frac{2\xi /D}{1+\sqrt{1+(2\xi /D)^{2}}}
\label{BinfDef}
\end{equation}
and is $B_{\infty }(\xi )$ the Langevin function for $D\gg 1$. Alternatively
Eq. (\ref{mvsM}) can be derived from Eq. (\ref{Mviam}) replacing $%
D-1\Rightarrow D$ and neglecting $dm/dx$ in the limit $D\rightarrow \infty $
and then solving the resulting algebraic equation for $m.$ One can find in
the literature formulae of the type $m=M_{s}B(\mathcal{N}M_{s}H/T)$, where
the saturation magnetization $M_{s}$ is usually associated with the bulk
magnetization at a given temperature (see, e.g., Refs.~\cite{fispri85}). In
our case, Eq.\ (\ref{mvsM}) is exact and $M=M(T,H)$ is defined by Eq.\ (\ref
{BinfDef}). For large sizes $\mathcal{N}$, Eq.\ (\ref{mvsM}) describes two
distinct field ranges separated at $H\sim H_{V}$ where
\begin{equation}
H_{V}\equiv \frac{TD}{\mathcal{N}M}.  \label{HVDef}
\end{equation}
In the range $H\lesssim H_{V}$ the total magnetic moment of the system is
disoriented by thermal fluctuations, $m<M$. In the range $H\gtrsim H_{V}$,
the total magnetic moment is oriented by the field, $m$ approaches $M$, and
both further increase with the field towards saturation ($m=M=1$) due to the
suppression of spin waves in the system. This scenario is inherent to all $%
O(D)$ models \cite{fispri85}.

Having established the superparamagnetic relation, Eq. (\ref{mvsM}) we are
left with the problem of calculating $M(T,H)$. For the pbc the solution
becomes homogeneous and one obtains (\ref{bulkEqs}) where in $P(G)$ the
integral is replaced by a sum over discrete wave vectors \cite{kacgar01physa}
, whereas $M=\sqrt{m^{2}+\theta G/\left[ \mathcal{N}(1-G)\right] }.$ For the
model with fbc analytical solution is only possible at low and high
temperatures. At $\theta \ll 1$ small deviations from the collinear state
with $M=1$ can be described by the modified SWT for arbitrary $D,$ see Sec. ~
\ref{sec:SWTMC}.
%
\begin{figure}[tbp]
\includegraphics[angle=-90,width=7cm]{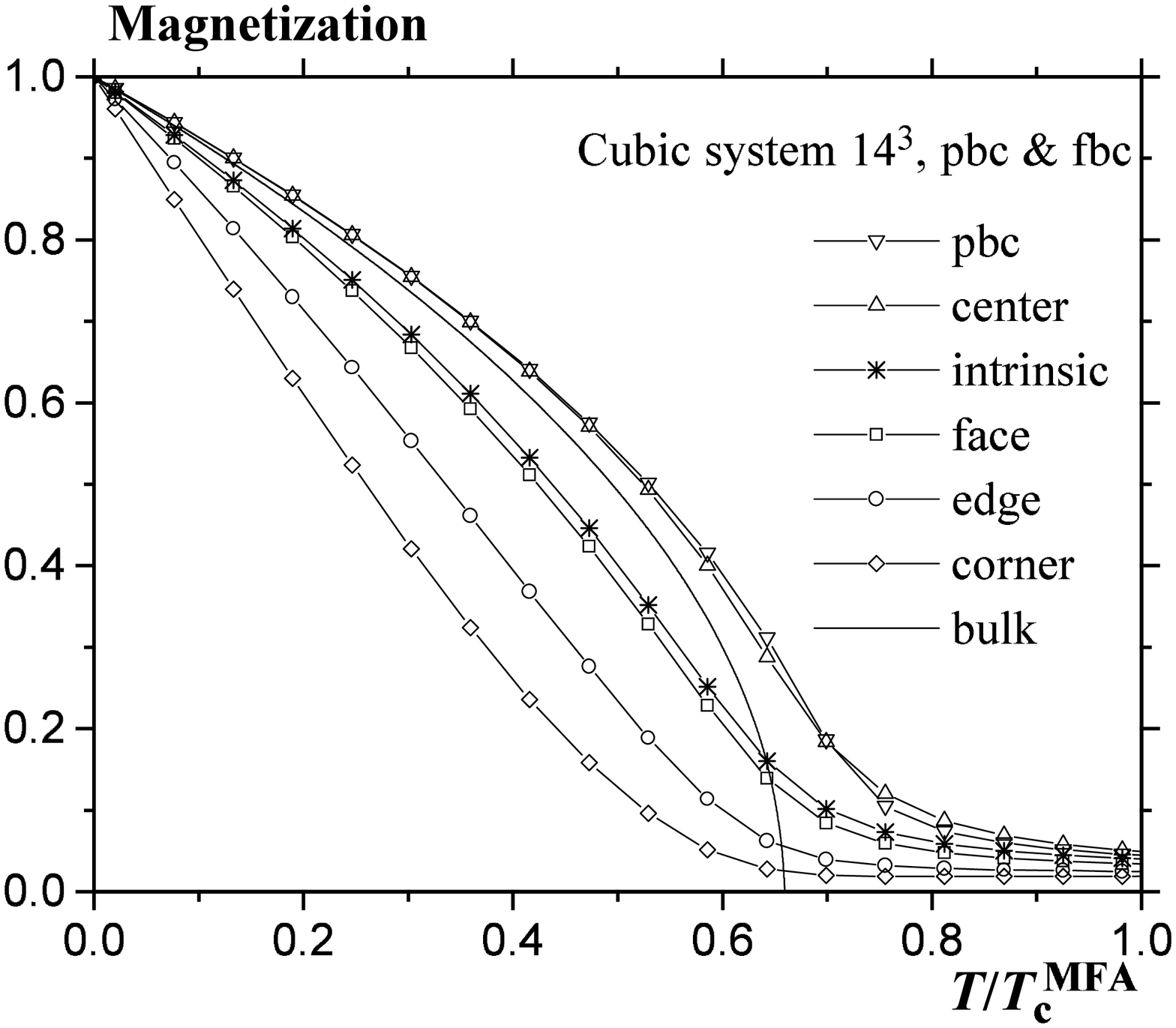} %
\includegraphics[angle=-90,width=7cm]{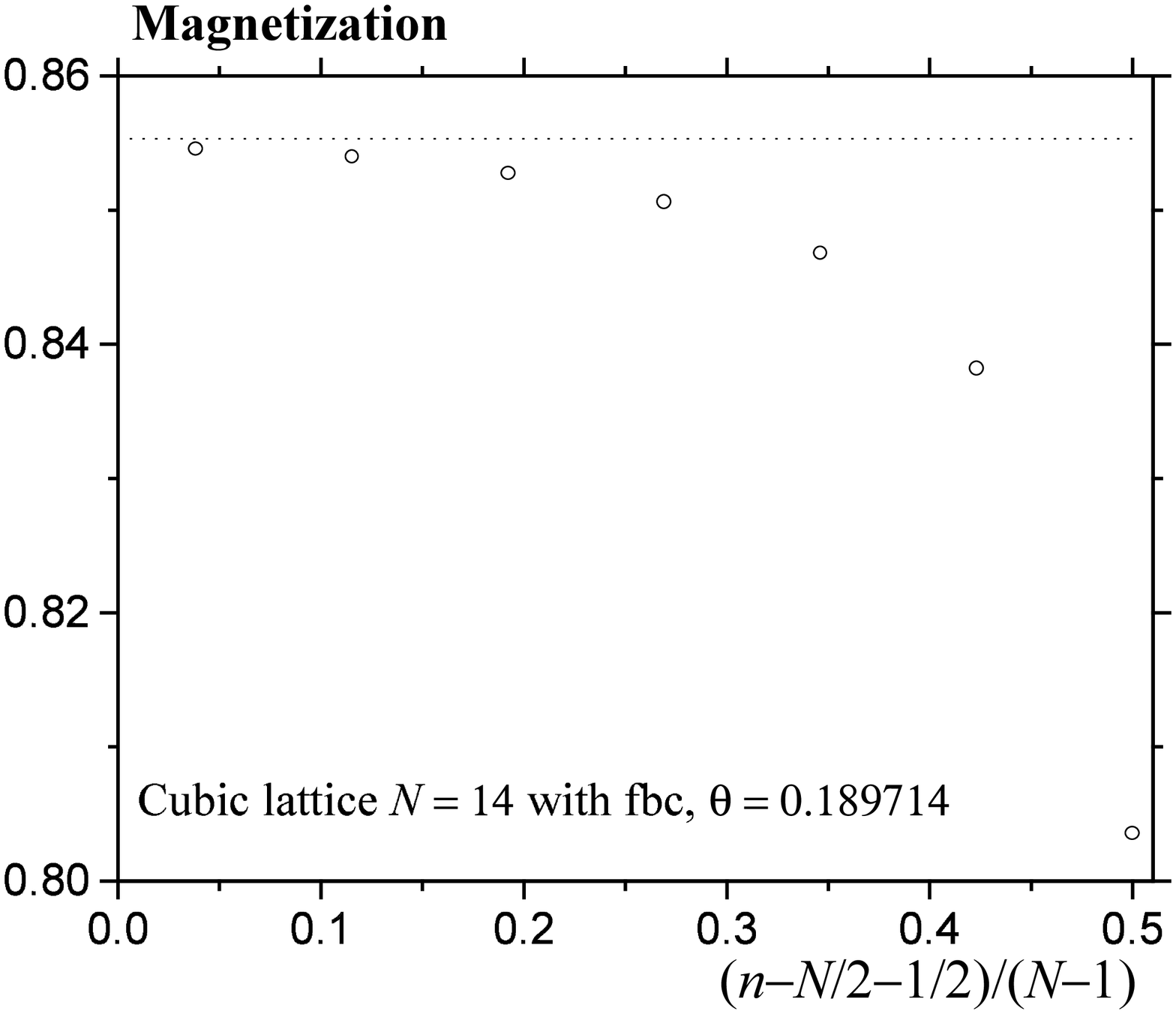}
\caption{Left: Temperature dependence of the intrinsic magnetization $M$,
Eq.\ (\ref{BinfDef}), and local magnetizations $M_{i}$, Eq.\ (\ref{MLocalDef}
) in zero field. Right: Long-range magnetization profile in the direction
from the center of the cube to the center of a face at temperature
$\protect\theta\equiv T/T_{c}^{\mathrm{MFA}} = 0.189714$.}
\label{sef_t3d}
\end{figure}
%
\subsubsection{ Numerical results and discussion}

The method for solving the $D\rightarrow \infty $ model consists in
obtaining the correlation functions $s_{ij}$ and magnetization $m_{i}$ from
the first two linear equations in Eq. (\ref{DefMatr}), substituting them
into the third equation of Eq. (\ref{DefMatr}), and solving the resulting
system of nonlinear equations for the gap parameter $G_{i}$ numerically.
Fig.\ \ref{sef_t3d} (left) shows the temperature dependence of the intrinsic
magnetization $M$, Eq.\ (\ref{BinfDef}), and local magnetizations $M_{i}$,
Eq.\ (\ref{MLocalDef}), of the $14^{3}$ cubic system with free and periodic
boundary conditions in zero field. For periodic boundary conditions, $M$
exceeds the bulk magnetization at all temperatures. In particular, at low
temperatures this is in accord with the positive sign of the finite-size
correction to the magnetization, Eqs.~(\ref{MLowTH0}) and (\ref{DeltaN}).
Local magnetizations at the center of the faces and edges and those at the
corners decrease with temperature much faster than the magnetization at the
center. One can see that below the bulk critical temperature $M$ is smaller
than the bulk magnetization. This means that the boundary effects
suppressing $M$ are stronger than the finite-size effects that increase $M$.
This is also seen from the low-temperature expression of $M$ given in Eq.~(%
\ref{MLowTH0}), see also Fig.~\ref{sef_del}. Fig.\ \ref{sef_t3d} (right) shows
the magnetization profile in the direction from the center of the cube to the
center of a face. It is seen that perturbations due to the free boundaries
extend deep into the particle, whereas the MFA predicts, on the contrary, a
fast approach to a constant magnetization when moving away from the boundary
\cite{wil74}. This is a consequence of the Goldstone mode which renders the
correlation length of an isotropic bulk magnet infinite below $T_{c}$. MC
simulations of the classical Heisenberg model \cite{wil74}, \cite
{kacetal00epjb} yield a similar result (see Fig.~\ref{Barce3} right).

%
\begin{figure}[tbp]
\centerline{\includegraphics[angle=-90,width=10cm]{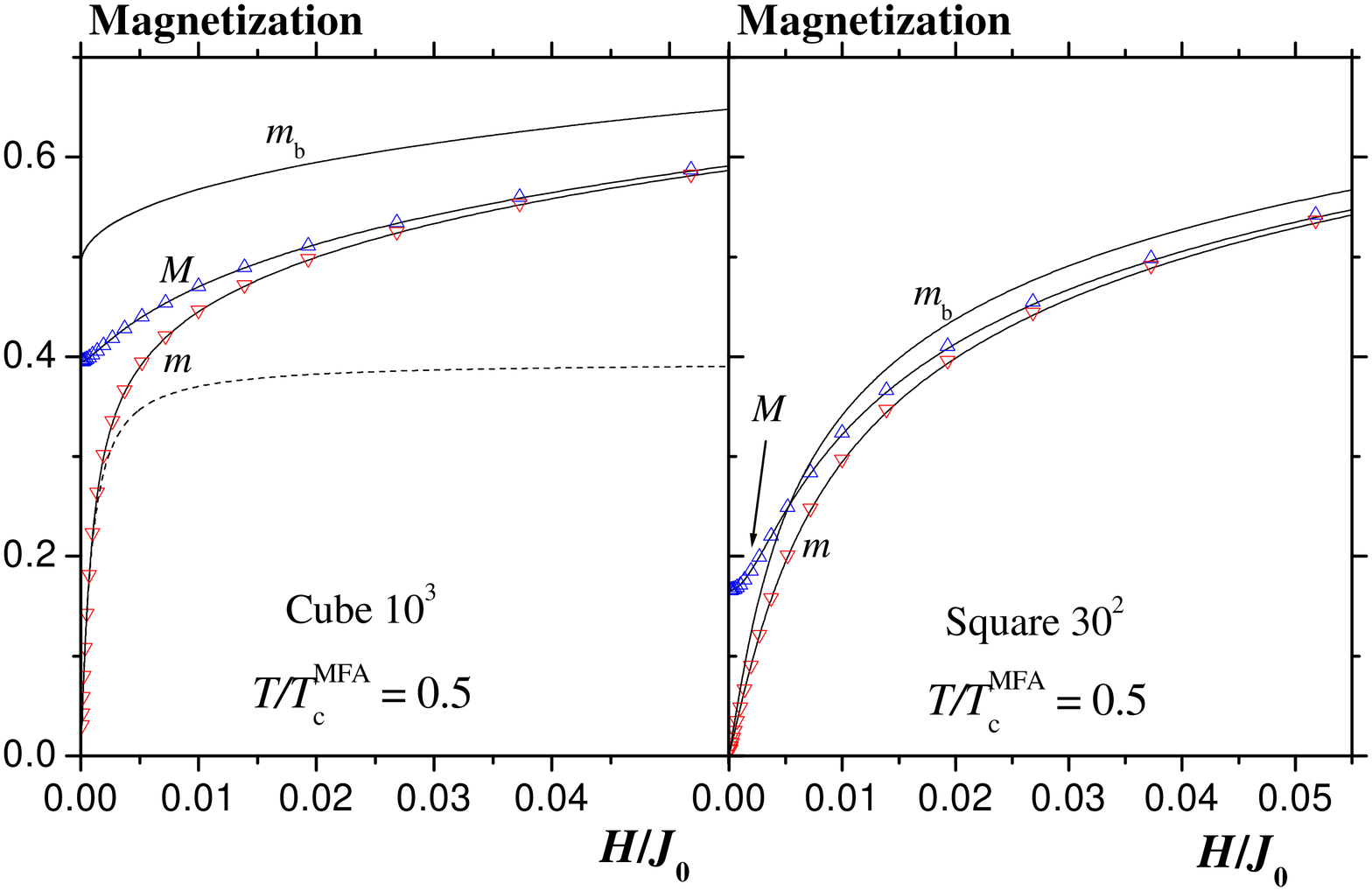}}
\caption{Field dependence of the intrinsic magnetization $M$ and induced
magnetization $m$ for hypercubic lattices with fbc in three and two
dimensions. Dashed line is a plot of Eq.~(\ref{mvsM}) in which $M(H,T)$ is
replaced by its zero-field value. Bulk magnetization $m_{\mathrm{b}}$ in two
and three dimensions is shown by solid lines. }
\label{sef_h}
\end{figure}
%

We have shown that the critical indices for the magnetization at the faces,
edges, and corners are higher than the bulk critical index $\beta =1/2$ for
the present $D=\infty $ model. The critical index at the face $\beta _{1}$
is the most studied surface critical index (see, for a review, Refs.~\cite
{bin83ptcp}). The exact solution of Bray and Moore \cite{bramoo77prl} for
the correlation functions at criticality in the $D=\infty $ model and
application of the scaling arguments yield the value $\beta _{1}=1$ (see
Table II in \cite{bin83ptcp}). Exact values of the edge and corner
magnetization indices, $\beta _{2}$ and $\beta _{3}$, seem to be unknown for
$D=\infty $. Cardy \cite{car83} used the first-order $\varepsilon $%
-expansion to obtain $\beta _{2}(\alpha )$ for the edge with an arbitrary
angle $\alpha $. For $\alpha =\pi /2$ and $D=\infty $ in three dimensions
the result for the edge critical exponent reads $\beta
_{2}=13/8+O(\varepsilon ^{2})=1.625+O(\varepsilon ^{2})$. To estimate the
magnetization critical indices in our model we have performed a
finite-size-scaling analysis (see, for a review, Ref. \cite{bin92fss})
assuming the scaling form $M=N^{-\beta /\nu }F_{M}(\tau N^{1/\nu })$ and
plotting the magnetization times $N^{\beta /\nu }$ vs $\tau N^{1/\nu }$.
Here $\nu =1$ is the critical index for the correlation length in the bulk
and $\tau \equiv T/T_{c}-1$, where $T_{c}=T_{c}^{\mathrm{MFA}}/W$ is the
bulk Curie temperature. Our results for the system with $N=10$ and $N=14$
merge into single ``master curves'' for $\beta _{1}=0.86$, $\beta _{2}=1.33$%
, and $\beta _{3}=1.79$, which have been obtained by fitting $M\propto
N^{-\beta /\nu }$ at $T=T_{c}$, i.e., $\theta =\theta _{c}=1/W$. Note that
our value 0.86 for the surface magnetization critical index $\beta _{1}$ is
substantially lower than the value $\beta _{1}=1$ following from scaling
arguments. This disagreement is probably due to corrections to scaling which
could be pronounced for our small linear sizes $N=10$ and $14$. A more
efficient way for obtaining an accurate value of $\beta _{1}$ is to perform
a similar analysis for the semi-infinite model. The latter was considered
analytically and numerically for $T\geq T_{c}$ and $H=0$ in \cite{gar98pre}.
We also mention the Monte Carlo simulations of the Ising model \cite
{plesel98} which yield $\beta _{1}=0.80$, $\beta _{2}=1.28$, and $\beta
_{3}=1.77$.

The field dependence of $M$ and $m$ at fixed temperature, as obtained from
the numerical solution of Eqs.~(\ref{DefMatr}) for cubic and square systems
is shown in Fig.~\ref{sef_h}. Naturally the numerical results for $m$
confirm Eq.~(\ref{mvsM}) which describes both the effect of orientation of
the system's magnetization by the field and the increase of $M$ in field.
Using the zero-field value of $M$ in Eq.~(\ref{mvsM}) leads to a poor result
for $m$ as shown by the dashed curve for the 10$^{3}$ system in Fig.~\ref
{sef_h}. The field dependence of the particle's magnetization similar to
that shown in Fig.~\ref{sef_h} for the cubic system was experimentally
obtained for ultrafine cobalt particles in \cite{resetal98prb}, as well as
in a number of previous experiments. The curves for the square system in
Fig.~\ref{sef_h} illustrate the fact that in two dimensions thermal
fluctuations are much stronger than in 3$d$, which leads to lower values of
both $M$ and $m$ at a given temperature. The bulk magnetization $m_{\mathrm{b
}}$ in two dimensions vanishes at zero field and it thus goes below the
intrinsic magnetization $M$ in the low-field region.

\subsection{\label{sec:SWTMC}Modified spin-wave theory: Low-temperature
properties and the superparamagnetic relation}

As briefly discussed in Sec.~\ref{sec:basic}, and in more details in the
previous section, it is important to investigate the precise relation
between the induced magnetization $\mathbf{m}$ of Eq. (\ref{mdef}) and
intrinsic magnetization $M$ of Eq. (\ref{Mdef}) for the more realistic
Heisenberg model, $D=3$. To do so, we have developed a finite-size spin-wave
theory that yields analytical results at low temperatures. We also performed
simulations with the improved Monte Carlo technique \cite{kacgar01epjb}.

\subsubsection{\label{sec:swtfsmp}Spin-wave theory for finite-size magnetic
particles}

In the absence of SA, at low temperatures all spins in the particle are
strongly correlated and form a ``giant spin'' $\mathbf{M}$ defined in Eq.~(%
\ref{MvecDef}) which behaves superparamagnetically. In addition, there are
internal spin-wave excitations in the particle that are responsible for $%
M(H,T)<1$ at $T>0$. These excitations can be described perturbatively in
small deviations of individual spins $\mathbf{s}_{i}$ from the global
direction of $\mathbf{M}$. Thus we use $\mathbf{M=}\mathcal{M}\mathbf{n}$
with $|\mathbf{n}|=1$ and insert an additional integration over $d\mathbf{M}=%
\mathcal{M}^{D-1}d\mathcal{M}d\mathbf{n}$ in the partition function,
\begin{equation}
\mathcal{Z}=\int \mathcal{M}^{D-1}d\mathcal{M}d\mathbf{n}\prod_{i}d\mathbf{s}%
_{i}\delta \left( \mathbf{M-}\frac{1}{\mathcal{N}}\sum_{i}\mathbf{s}%
_{i}\right) \mathrm{e}^{-\mathcal{H}/T},  \label{PartFunc}
\end{equation}
and first integrate over the magnetization magnitude $\mathcal{M}$. Thus we
reexpress the vector argument of the $\delta $-function in the coordinate
system specified by the direction of the central spin $\mathbf{n}:$
\begin{equation}
\delta \left( \mathbf{M}-\frac{1}{\mathcal{N}}\sum_{i}\mathbf{s}_{i}\right)
=\delta \left( \mathcal{M}-\frac{1}{\mathcal{N}}\sum_{i}(\mathbf{n\cdot s}%
_{i})\right) \delta \left( \frac{1}{\mathcal{N}}\sum_{i}\left[ \mathbf{s}%
_{i}-\mathbf{n}(\mathbf{n\cdot s}_{i})\right] \right) .
\end{equation}
Then after integration over $\mathcal{M}$ one obtains
\begin{equation}
\mathcal{Z}=\int d\mathbf{n}\mathcal{Z}_{\mathbf{n}},\qquad \mathcal{Z}_{%
\mathbf{n}}=\int \prod_{i}d\mathbf{s}_{i}\delta \left( \frac{1}{\mathcal{N}}%
\sum_{i}\left[ \mathbf{s}_{i}-\mathbf{n}(\mathbf{n\cdot s}_{i})\right]
\right) \mathrm{e}^{-\mathcal{H}_{\mathrm{eff}}/T},  \label{Zndef}
\end{equation}
where $\mathcal{Z}_{\mathbf{n}}$ is the partition function for the fixed
direction $\mathbf{n}$ and
\begin{equation}
\mathcal{H}_{\mathrm{eff}}=-(\mathbf{n\cdot H})\sum_{i}(\mathbf{n\cdot s}%
_{i})-\frac{1}{2}\sum_{ij}J_{ij}\mathbf{s}_{i}\cdot \mathbf{s}_{j}-(D-1)T\ln %
\left[ \frac{1}{\mathcal{N}}\sum_{i}\mathbf{n\cdot s}_{i}\right] .
\label{Heff}
\end{equation}
In Eq.~(\ref{Zndef}), the $\delta $-function says that the sum of all spins
does not have a component perpendicular to $\mathbf{M}$. This will lead to
the absence of the $\mathbf{k}=0$ component of the transverse spin
fluctuations. That is, the global-rotation Goldstone mode that is
troublesome in the standard spin-wave theory for finite systems, has been
transformed into the integration over the global direction $\mathbf{n}$ in
Eq.~(\ref{Zndef}). $\mathcal{Z}_{\mathbf{n}}$ is computed at low temperature
by expanding $\mathcal{H}_{\mathrm{eff}}$ up to bilinear terms in the
transverse spin components $\mathbf{\Pi }_{i}=\mathbf{s}_{i}-\mathbf{n}(%
\mathbf{n\cdot s}_{i})$:
\begin{eqnarray}
\mathcal{H}_{\mathrm{eff}} &\cong &E_{0}-\mathcal{N}\mathbf{n\cdot H}+\frac{1%
}{2}\sum_{ij}A_{ij}\mathbf{\Pi }_{i}\cdot \mathbf{\Pi }_{j},  \nonumber
\label{HamPtrans} \\
A_{ij} &\equiv &\left[ (D-1)T/\mathcal{N}+\mathbf{n\cdot H+}\sum_{l}J_{il}%
\right] \delta _{ij}-J_{ij},
\end{eqnarray}
where $E_{0}=-(1/2)\sum_{ij}J_{ij}$ is the zero-field ground-state energy.
Next, upon computing the resulting Gaussian integrals over $\Pi _{i}^{\alpha
}$, one obtains
\begin{equation}
\mathcal{Z}_{\mathbf{n}}\cong \mathrm{\exp }\left( \frac{-E_{0}+\mathcal{N}%
\mathbf{n\cdot H}}{T}\right) \mathcal{N}^{D-1}\left[ \frac{(2\pi T)^{%
\mathcal{N}-1}}{\prod_{k}{}^{^{\prime }}\frac{A_{k}}{\mathcal{N}}}\right]
^{(D-1)/2},  \label{ZnDeterm}
\end{equation}
where for particles of cubic shape
\begin{equation}
A_{\mathbf{k}}=A_{0}+J_{\mathbf{k}}-J_{0},\qquad A_{0}=\sum_{i}A_{ij}=(D-1)T/%
\mathcal{N}+\mathbf{n\cdot H.}  \label{AkDef}
\end{equation}
The prime on the product in (\ref{ZnDeterm}) means omitting the $k=0$ mode.
Results for particles of arbitrary shape can be found in Ref. \cite
{kacgar01epjb}. In Eq. (\ref{ZnDeterm}) the exponential factor corresponds to
rigid spins whereas the second factor describes spin-wave corrections. The
latter makes the angular dependence of $\mathcal{Z}_{\mathbf{n}}$ more
complicated. Differentiating $\mathcal{Z}$ with respect to $H$ yields the
induced magnetization $m,$ then the intrinsic magnetization $M$ can be
obtained from the exact relation (\ref{Mviam}), and the validity of the
superparamagnetic relation (\ref{spmrelation}) can be checked.

\subsubsection{\label{mMspmr}Induced and intrinsic magnetizations, and
superparamagnetic relation}

Now we consider particles of cubic shape ($\mathcal{N}=N^{3}$) at low
temperatures. For both pbc and fbc, we have obtained the following
correction to the magnetization in zero field \cite
{kacgar01physa,kacgar01epjb}
\begin{equation}
M\cong 1-t,\qquad t\equiv \frac{D-1}{2}\frac{W_{N}T}{J_{0}} =
\frac{D-1}{2D}W_N\theta,\qquad
W_{N}=\frac{1}{\mathcal{N}}\sum_{\mathbf{k}}{}^{^{\prime }}\frac{1}{1-\lambda
_{\mathbf{k}}}  \label{MLowTH0} \end{equation}
where $W_{N}$ is the sum without the $\mathbf{k}=0$ term. The results for
pbc and fbc differ only by the values of the discrete wave vectors in Eq. (%
\ref{MLowTH0}) \cite{kacgar01physa}:
\begin{equation}
k_{\alpha }=\left\{
\begin{array}{cc}
2\pi n_{\alpha }/N, & \mathrm{pbc} \\
\pi n_{\alpha }/N, & \mathrm{fbc}
\end{array}
\right. ,\qquad n_{\alpha }=0,1,...,N-1  \label{defkpbc}
\end{equation}
where $\alpha =x,y,z.$ This subtle difference is responsible for much
stronger thermal fluctuations in the fbc model due to boundary effects. The
limit $N\rightarrow \infty $ of $W_{N}$ is the so-called Watson's integral $%
W=P(0)$ of Eq. (\ref{bulkEqs}). The difference between $W_{N}$ and $W$ has
different signs for pbc and fbc models \cite{kacgar01physa}
\begin{equation}
\Delta _{N}\equiv \frac{W_{N}-W}{W}\cong \left\{
\begin{array}{ll}
\displaystyle-\frac{0.90}{N}, & \mathrm{pbc} \\
\displaystyle\frac{9\ln (1.17N)}{2\pi WN}, & \mathrm{fbc.}
\end{array}
\right.  \label{DeltaN}
\end{equation}
%
\begin{figure}[tbp]
\centerline{\includegraphics[angle=-90,width=8cm]{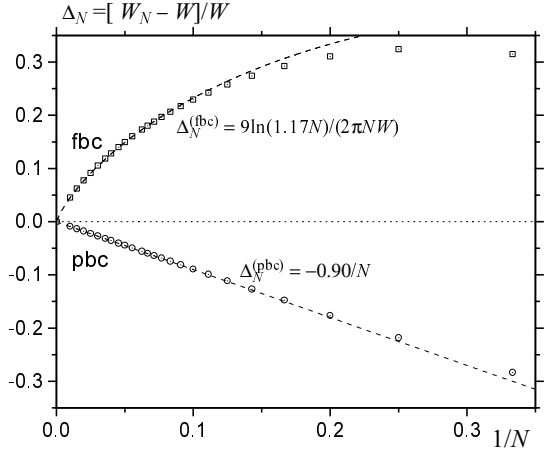}}
\caption{Lattice sums $W_{N}$ for cubic systems with free and periodic
boundary conditions. $W=1.51639$ is the bulk value for the sc lattice.}
\label{sef_del}
\end{figure}
%
Therefore, Fig.~\ref{sef_del} shows that the coefficient in the linear-$%
\theta $ term in Eq.~(\ref{MLowTH0}) is smaller than in the bulk for the pbc
system and greater for the fbc system. That is, boundary effects suppress
the intrinsic magnetization at low temperatures while finite-size effects
lead to its increase. Fig.~\ref{sef_del} shows that boundary effects render
a larger contribution than the finite-size effects, making the net
magnetization well below that of the bulk.

The field dependence of $m$ and $M$ is defined by the expansion of the
lattice Green function for small gaps, $1-G\ll 1$
\begin{equation}
\widetilde{P}_{N}(G)=\frac{1}{\mathcal{N}}\sum_{\mathbf{k}}{}^{^{\prime }}%
\frac{1}{1-G\lambda _{\mathbf{k}}}\cong W_{N}-\left\{
\begin{array}{cc}
c_{N}N(1-G), & N^{2}(1-G)\ll 1 \\
c_{0}\sqrt{1-G}, & N^{2}(1-G)\gg 1,
\end{array}
\right.  \label{PGtilExp}
\end{equation}
where for the sc lattice $c_{0}=(2/\pi )(3/2)^{3/2}$ and the numerical
results for $c_{N}$ can, for $N\gg 1$, be fitted as
\begin{equation}
c_{N}\cong \left\{
\begin{array}{ll}
0.384-1.05/N, & \mathrm{pbc} \\
1.90-1.17/N, & \mathrm{fbc.}
\end{array}
\right.  \label{CN}
\end{equation}
The spin-wave gap $1-G$ \ depends on the temperature and field; For $H\ll
H_{V}$ [see Eq. (\ref{HVDef})] the gap approaches its zero-field value while
for $H\gg H_{V}$ one has $1-G\cong h\equiv H/J_{0}.$ Thus Eq. (\ref{PGtilExp}%
) defines one more crossover in field, the crossover between its first and
second lines at
\begin{equation}
H_{S}\sim \frac{J_{0}}{\mathcal{N}^{2/3}}=\frac{J_{0}}{N^{2}}\gg H_{V}=\frac{%
TD}{\mathcal{N}M}=\frac{\theta J_{0}}{\mathcal{N}M}.  \label{HSDef}
\end{equation}
In the field range $H\gg H_{S}$ the discreteness of the lattice can be
neglected and the bulk result $\Delta M\sim \sqrt{H}$ is reproduced. In the
most interesting region $H\ll H_{S}$ one obtains
\begin{equation}
M\cong 1-t+2\alpha xB(x),\qquad \alpha \equiv \frac{(D-1)c_{N}}{4N^{2}}%
\left( \frac{T}{J_{0}}\right) ^{2}.  \label{Mresult}
\end{equation}
On the other hand, Eq.~(\ref{Mresult}) describes a crossover from the
quadratic field dependence of $M$ at low field, $x\ll 1,$ to the linear
dependence at $x\gg 1.$ Note that for $x\gg 1,$ where $m\cong M$ and a rigid
magnetic moment would saturate, $m$ continues to increase linearly as $%
m\cong 1-t+2\alpha x.$ This is due to the field dependence of the intrinsic
magnetization $M$. At higher fields there is another crossover to the
standard spin-wave theory expression for $M$. Approximate expressions for $M$
in the different field ranges are
\begin{equation}
M\cong 1-t+\left\{
\begin{array}{ll}
\displaystyle\frac{D-1}{2D}c_{N}\left( \frac{HN^{2}}{J_{0}}\right) ^{2}, &
H\ll H_{V} \\
\displaystyle\frac{D-1}{2}c_{N}\frac{NHT}{J_{0}^{2}}, & H_{V}\ll H\ll H_{S}
\\
\displaystyle\frac{D-1}{2}c_{0}\frac{T}{J_{0}}\left( \frac{H}{J_{0}}\right)
^{1/2}, & H_{S}\ll H\ll J_{0}.
\end{array}
\right.  \label{Mfinalres}
\end{equation}

A simple analysis shows \cite{kacgar01epjb} that the superparamagnetic relation
(\ref {spmrelation}) is a very good approximation for not too small systems, $
\mathcal{N}\gg 1$ in the whole range below $T_{c}.$ The deviation from Eq. (
\ref{spmrelation}) is controlled by the small parameter $\alpha $ of Eq. (
\ref{Mresult}). Above $T_{c},$ however, deviations from Eq. (\ref
{spmrelation}) are large, except for the model with $D\rightarrow \infty .$
In the close vicinity of $T_{c}$, there is a crossover to the high-temperature
form of Eq.~(\ref{spmrelation}) given by the function $B_{\infty }(x)$ of Eq. (
\ref{BinfDef}).

The modified SWT developed above can be applied to study inhomogeneities in
the fbc model. The local \emph{intrinsic} magnetization defined by Eq.~(\ref
{MLocalDef}) shows stronger temperature dependence near the boundary than
the averaged $M\cong 1-t$ of Eq. (\ref{MLowTH0}). The biggest effect of the
surface is naturally attained at the corners of the cube where $M_{i}\cong
1-8t$, at $H=0$ \cite{kacgar01epjb}.

\subsubsection{MC simulations}

Here we apply our Monte Carlo technique that accounts for the
global-rotation Goldstone mode to compute the induced and intrinsic
magnetizations, and to investigate the superparamagnetic relation between
them, for the Heisenberg model with fbc. Our results for Ising model can be
found in \cite{kacgar01epjb}.

\begin{figure}[t]
\centerline{\includegraphics[angle=-90,width=10cm]{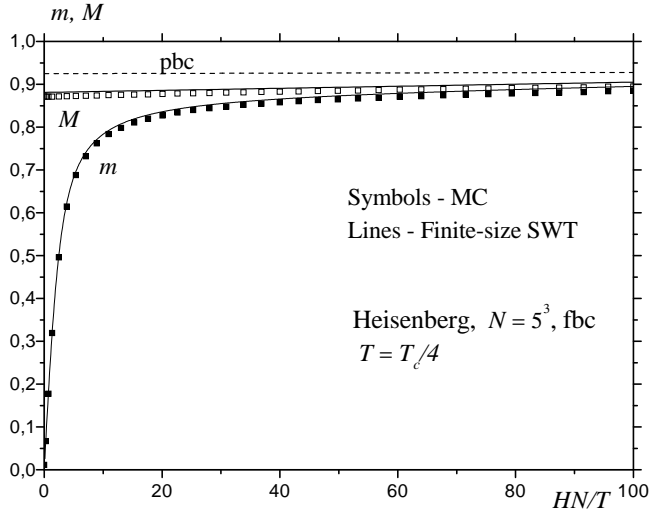}}
\caption{Comparison of the theoretical and MC results for the field
dependences of the magnetizations $M$ and $m$ for the Heisenberg model at $%
T=T_{c}/4$.}
\label{fig-hth}
\end{figure}

First of all, in Fig.~\ref{fig-hth} we compare theoretical predictions of
our analytical calculations within spin-wave theory for the Heisenberg model
with our MC results at $T=T_{c}/4,$ where $T_{c}=0.722T_{c}^{\mathrm{MFA}}$
is the actual bulk Curie temperature. %
%
For the small size $\mathcal{N}=5^{3}$ the square-root field dependence of
the magnetization (third line of Eq.~(\ref{Mfinalres})) does not arise and
finite-size corrections are very important. For $M$ one should use Eq.~(\ref
{Mresult}) with $t$ given by Eq.\negthinspace\ (\ref{MLowTH0}) with
numerically computed $W_{N}=1.99$ and $c_{N}=1.66$ for the fbc model [cf.
Eqs.~(\ref{DeltaN}) and (\ref{CN})]. This yields $t\simeq 0.119$ and $\alpha
\simeq 1.20\times 10^{-4}$. The theoretical dependence $M(H)$ is practically
a straight line which goes slightly above the MC points. This small
discrepancy can be explained by the fact that the applicability criterion
for our analytical method, $t\ll 1,$ is not fully satisfied at $T=T_{c}/4$%
, and a better agreement is achieved at lower temperatures. For comparison
we also plot the theoretical $M(H)$ for the model with periodic boundary
conditions. Here one has $W_{N}=1.25$ and $c_{N}=0.20$, thus $t\simeq 0.075$
and $\alpha \simeq 1.45\times 10^{-5}$, so $M(H)$ goes noticeably higher and
with a much smaller slope. The quadratic field dependence of $M$ in the
region $x\lesssim 1$ is not seen at this low temperature since the value of $%
\alpha $ is very small and thus much more accurate MC simulations would be
needed. We also plot in Fig.~\ref{fig-hth} the analytical result for the
field dependence of $m$ \cite{kacgar01epjb} which favorably compares with our MC
data.
%
\begin{figure}[t]
\includegraphics[angle=-90,width=7cm]{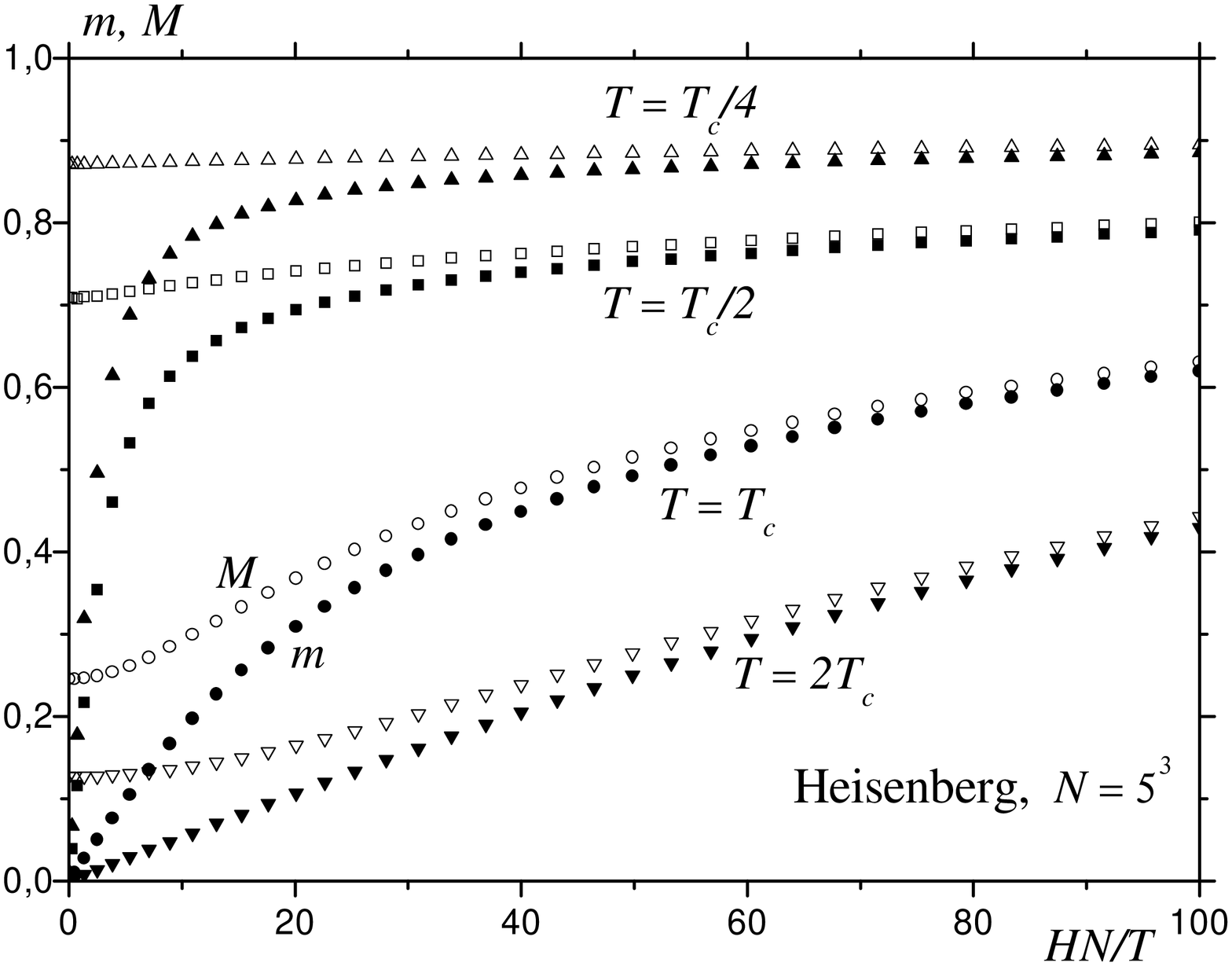} %
\includegraphics[angle=-90,width=7cm]{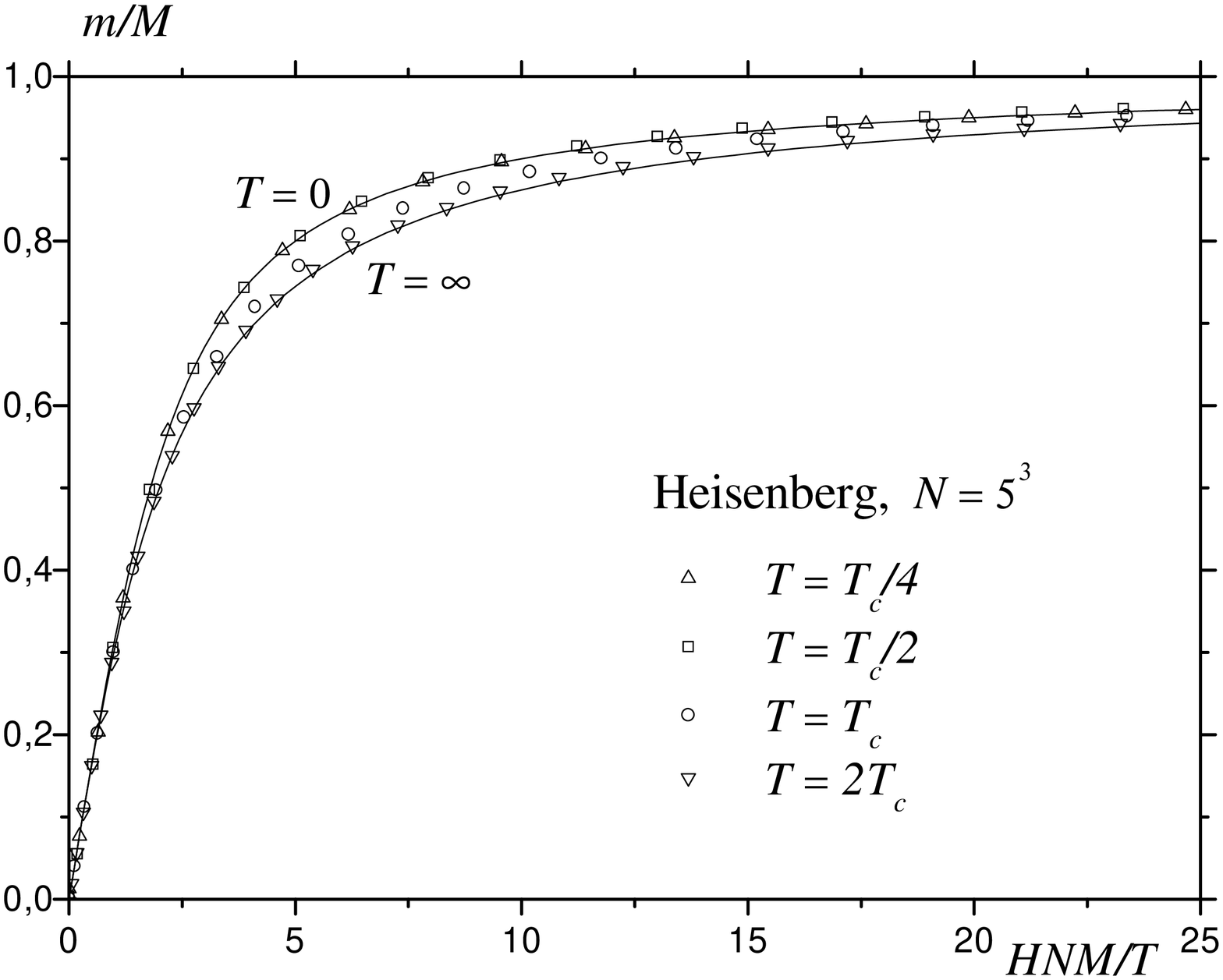}
\caption{Left: Field dependence of the intrinsic magnetization $M$ and the
induced magnetization $m$ of the Heisenberg model on the sc lattice with fbc
for different temperatures. Right: Scaled graph for $m/M$. Theoretical
curves $B_{3}(x)=\coth x-1/x$ for $T\ll T_{c}$ and $B_{\infty }(x)$ for $%
T\gg T_{c}$ are shown by solid lines. }
\label{fig-h}
\end{figure}
%
Fig.~\ref{fig-h} (left) shows the intrinsic magnetization $M$ and induced
magnetization $m$ versus the scaled field $x\equiv \mathcal{N}H/T$ for
different temperatures. We see that the particle's magnetic moment is
aligned and thus $m\sim M$ for $x\gtrsim 1$, if $T\ll T_{c}$. At $T\gg T_{c}$
the field aligns individual spins, and this requires $H\gtrsim T$, i.e., $%
x\gtrsim \mathcal{N}$. The quadratic dependence of $M(H)$ at small fields
manifests itself strongly at elevated temperatures.

The results of Fig.~\ref{fig-h} (right) show that the superparamagnetic
relation of Eq.~(\ref{spmrelation}) with $M=M(T,H)$ is a very good
approximation everywhere below $T_{c}$, for the Heisenberg model. This also
holds for the Ising model \cite{kacgar01epjb}. On the other hand, above $%
T_{c}$ Eq.~(\ref{spmrelation}) with the function $B_{\infty }(x)$ of Eq.~(%
\ref{BinfDef}) is obeyed. The difference between these limiting expressions
decreases with increasing number $D$ of spin components and disappears in
the spherical limit ($D\rightarrow \infty $).

\section{\label{conclusion}\label{sec:mspwse} A nanoparticle as a multi-spin
system: Effect of surface anisotropy}

Surface anisotropy causes large deviations from the bulk behavior that are
much stronger than just the effect of free boundaries. Also SA exerts
influence upon the coercive field. Here we first consider magnetic
structures and hysteresis loops at zero temperature induced by a strong
transverse surface anisotropy (TSA). Then for the more physically plausible
N\'{e}el's surface anisotropy (NSA) we investigate analytically and
numerically its contribution to the effective anisotropic energy of the
particle in the case when the NSA is weak in comparison to the exchange.
After that we study the effect of the NSA at finite temperatures using the
MC technique.
In the end of this section we investigate the thermal and spatial dependence of
the magnetization of a maghemite particle.

\subsection{\label{sec:LLEKs}Magnetic structure and hysteresis at $T=0$:
Transverse surface anisotropy}

Here we study the effect of TSA on the hysteresis loop and the angular
dependence of the switching field. We construct an effective
Stoner-Wohlfarth (SW) astroid for a single-domain spherical particle with
free surfaces, a simple cubic (sc) crystal structure, ferromagnetic exchange
$J,$ uniaxial anisotropy in the core $K_{c}$, and the TSA of strength $K_{s}$%
. Using the Hamiltonian (\ref{hamgeneral}) without the DDI at $T=0,$ we
solve the coupled Landau-Lifshitz equations (LLE) for each spin in the
particle~\cite{kacdim02prbjap} until a stationary state is attained. \ In
\cite{dimwys94prb} the same method was used for studying hysteresis loops in
nanoparticles with a random bulk or surface-only anisotropy. Here we address the
question of whether one can still use the simple SW model for a nanoparticle
endowed with strong surface effects. We show that it is so as long as $
K_{s}\lesssim J.$ Otherwise switching of the particle's magnetization occurs
via the reversal of clusters of spins, invalidating the simple SW model.

We consider $K_{s}$ and exchange coupling on the surface as free parameters
since there are so far no definite experimental estimations of them, whereas
the core parameters are taken as for the bulk system. In the sequel, we use
the reduced parameters, $j\equiv J/K_{c},k_{s}\equiv K_{s}/K_{c}$. %
\begin{figure}[tbp]
\centerline{\includegraphics[angle=-90,width=14cm]{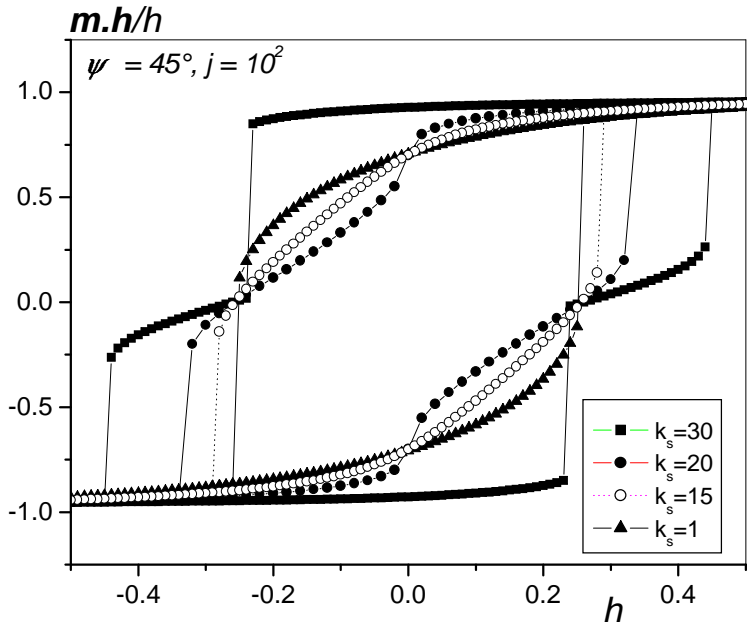}}
\caption{Hysteresis for $\protect\psi =\protect\pi /4$, $j=10^{2}$, $N=10$ $(%
\mathcal{N}=360)$ and different $k_{s}$. }
\label{hyst_ks_a45}
\end{figure}
%

\begin{figure}[h]
\centerline{\includegraphics[angle=-90,width=2.5cm]{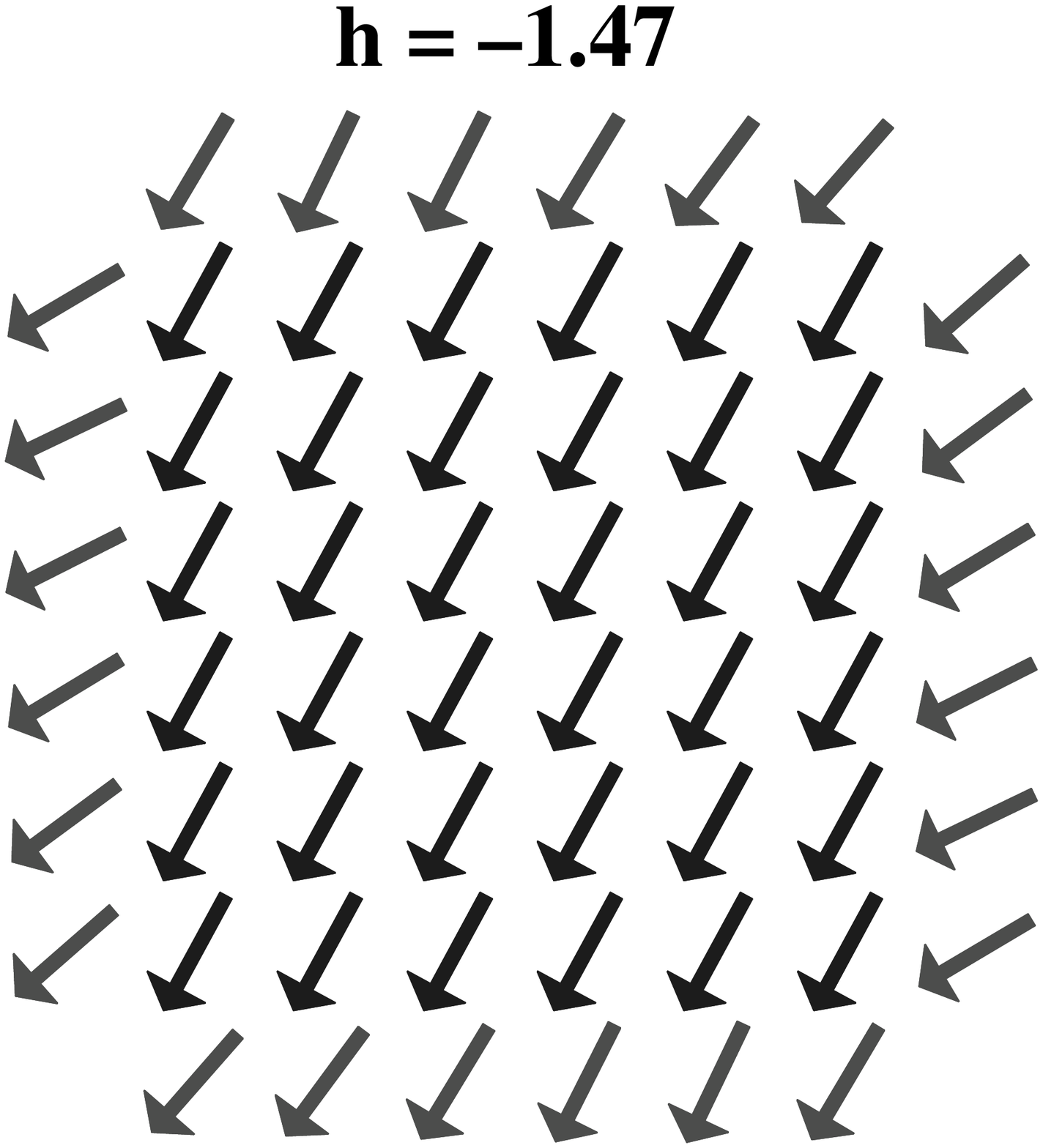}\hspace{1cm}
\includegraphics[angle=-90,width=2.5cm]{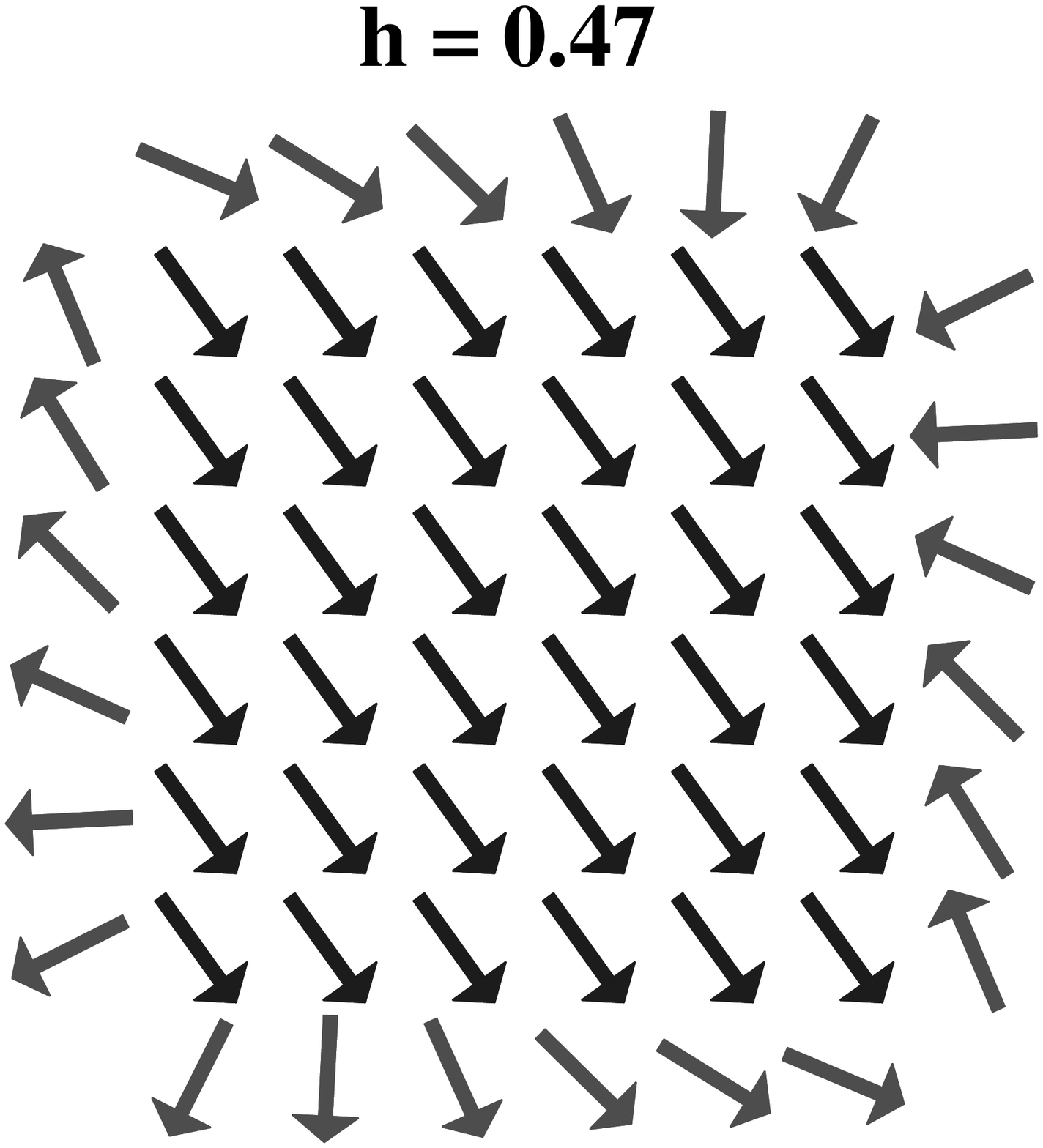}\hspace{1cm}
\includegraphics[angle=-90,width=2.5cm]{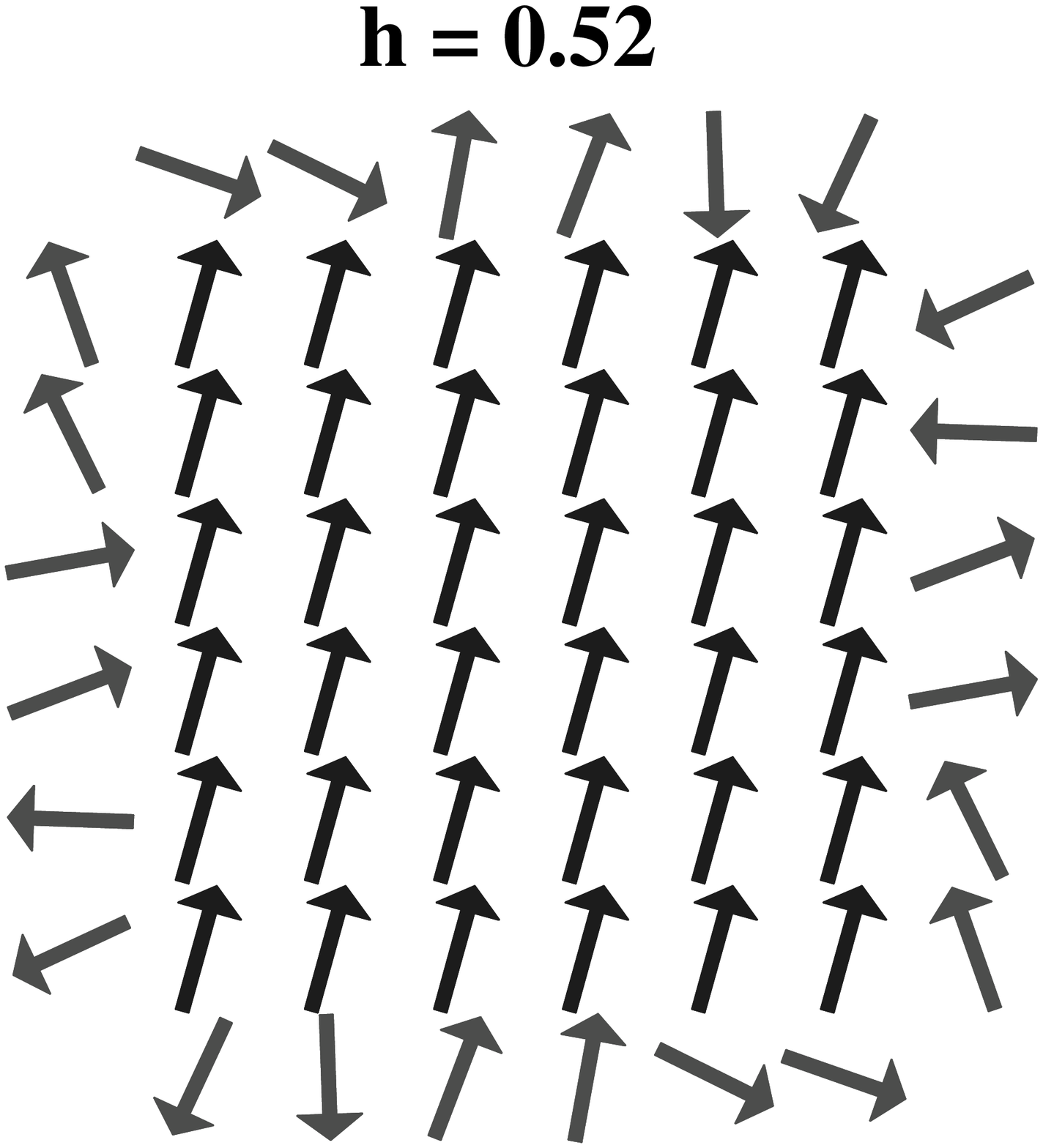}}
\centerline{\includegraphics[angle=-90,width=2.5cm]{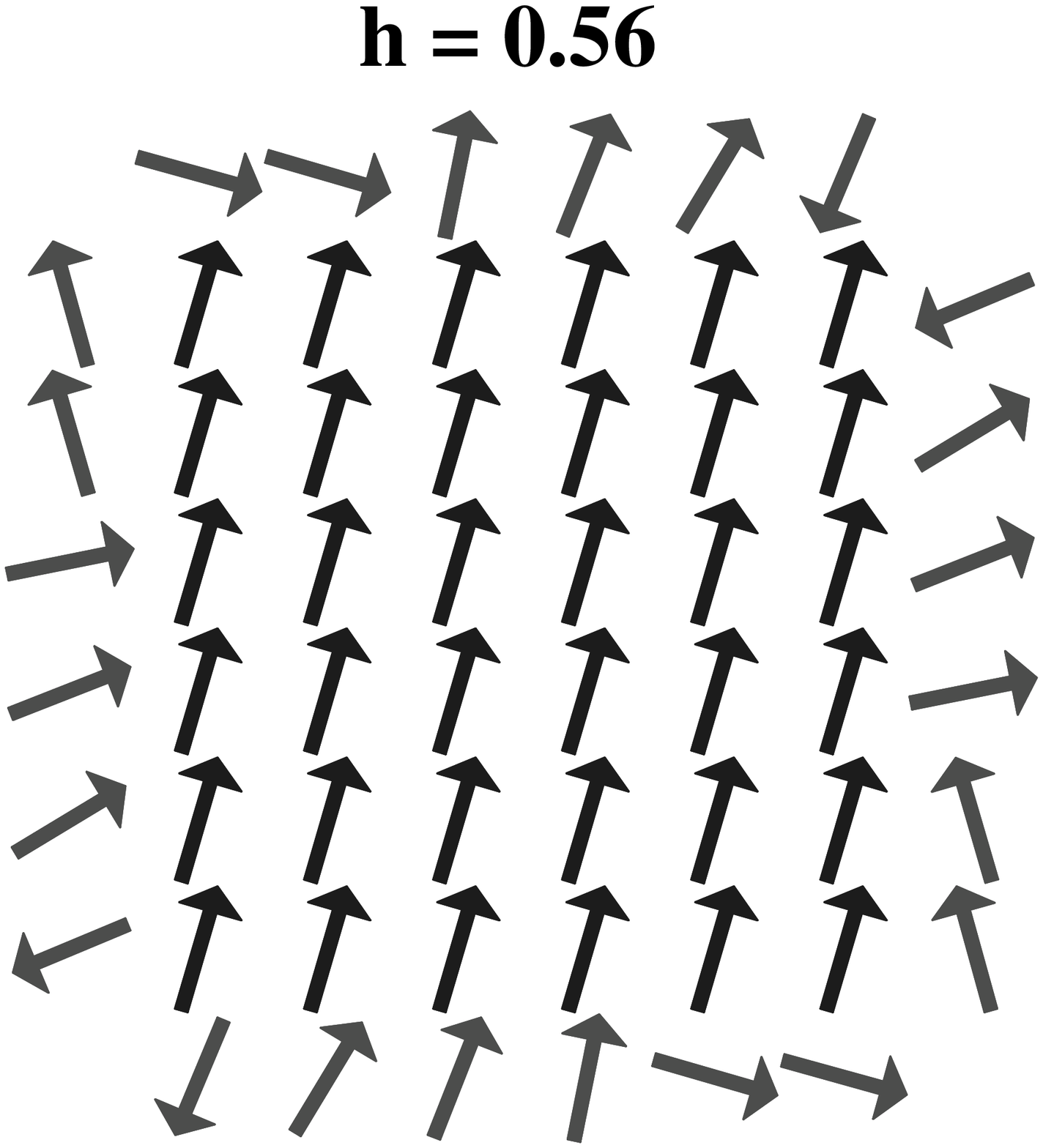}
\hspace{1cm} \includegraphics[angle=-90,width=2.5cm]{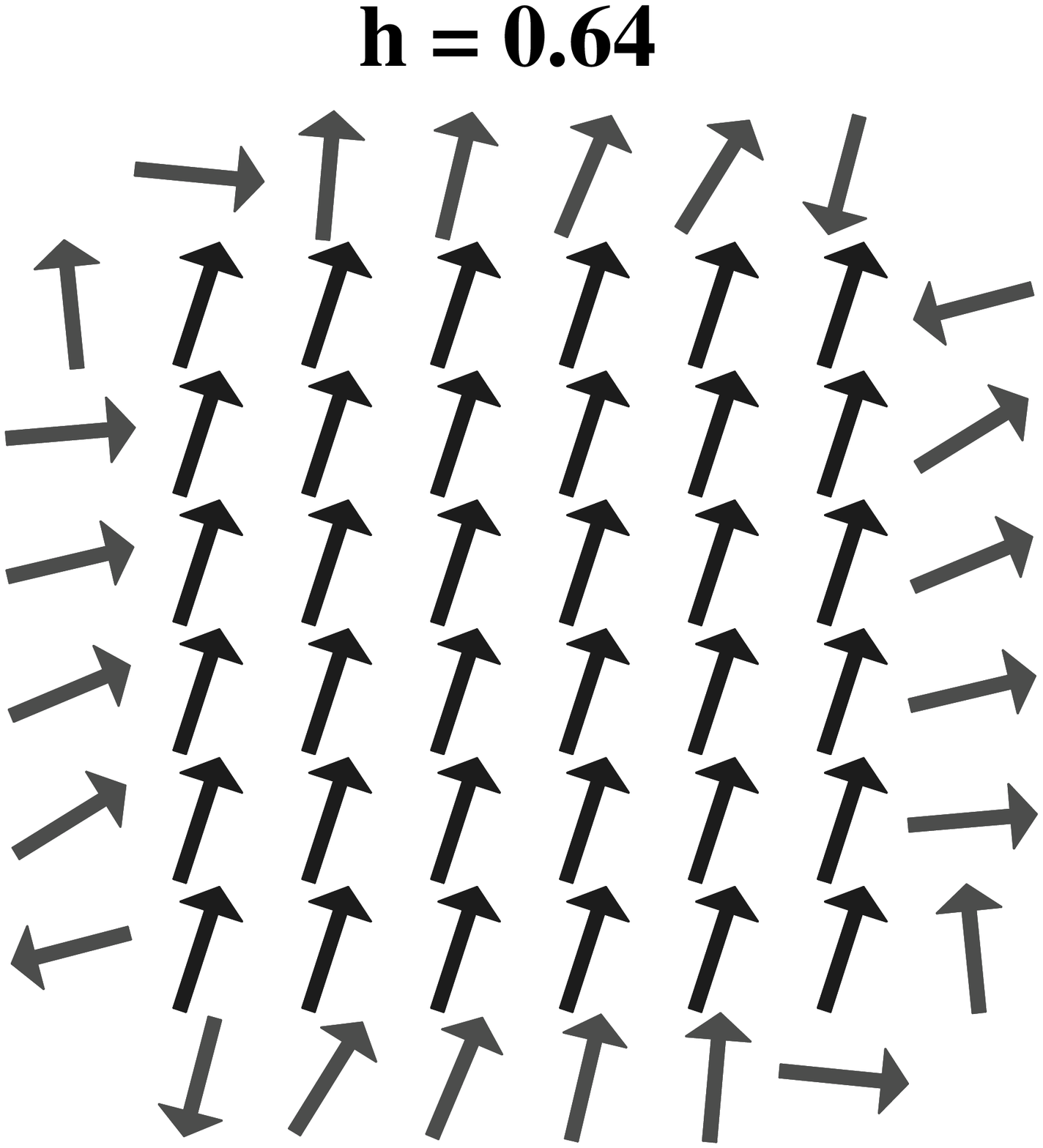}
\hspace{1cm}
\includegraphics[angle=-90,width=2.5cm]{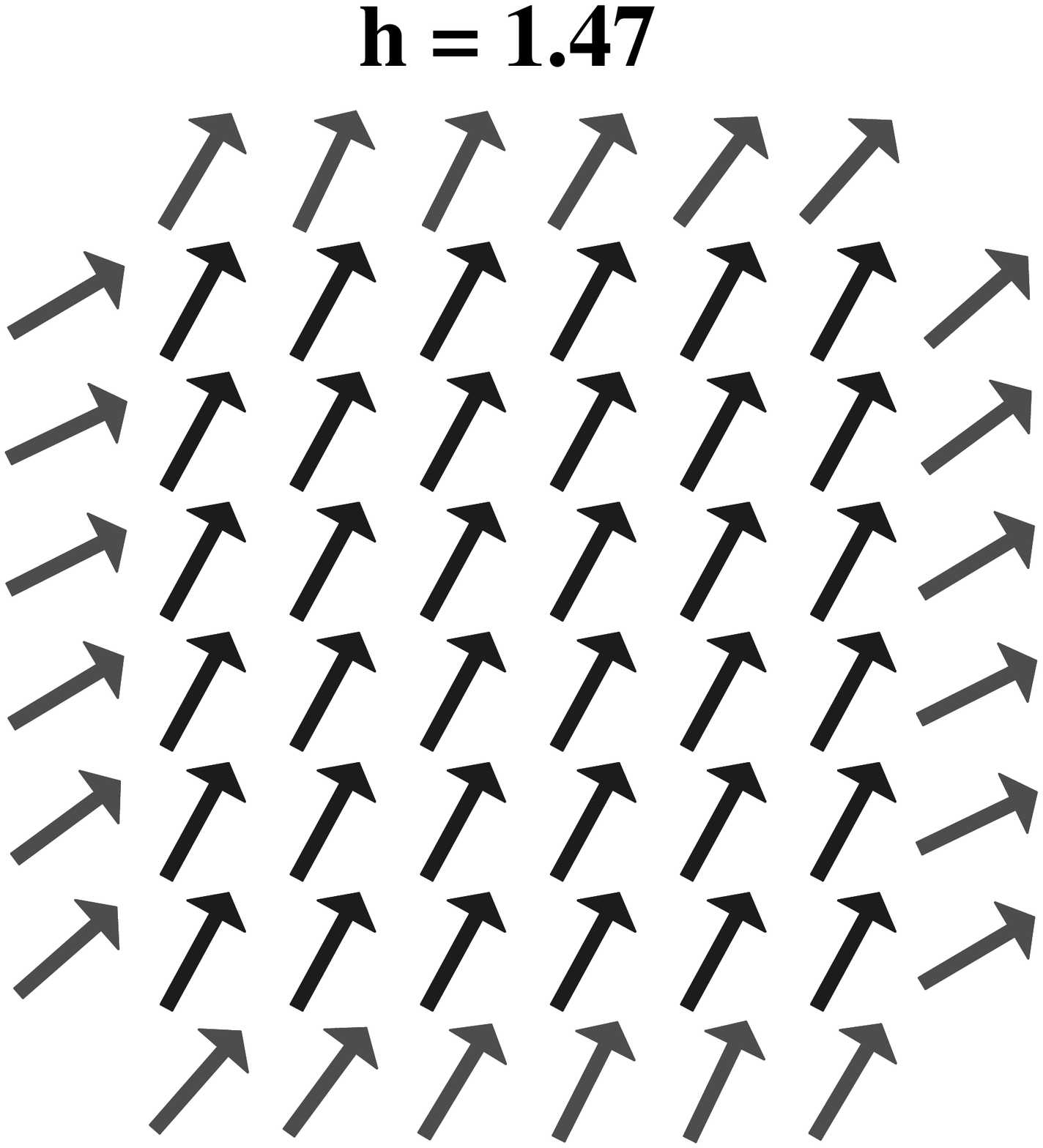}}
\caption{Magnetic structures in the middle plane of the particle with $%
\mathcal{N}=360$ in the TSA model for $j=0$, $k_{s}=1$, $\protect\psi =%
\protect\pi /4$ and different fields.}
\label{struct_T_D10_j0_K1}
\end{figure}

Fig.~\ref{hyst_ks_a45} shows that when $k_{s}$ becomes comparable with $j$,
the hysteresis loop exhibits multiple jumps, which can be attributed to the
switching of different spin clusters containing surface spins whose easy
axes make the same angle with the field direction. The hysteresis loop is
characterized by two field values: One that marks the limit of
metastability, called the \textit{critical field} or the saturation field,
and the other that marks the magnetization switching, and is called the
\textit{switching field} or the coercive field. This progressive switching
of spins is illustrated in Fig.~\ref{struct_T_D10_j0_K1} for simplicity for
the non-interacting case. %
%

\begin{figure}[tbp]
\includegraphics[angle=-90,width=14cm]{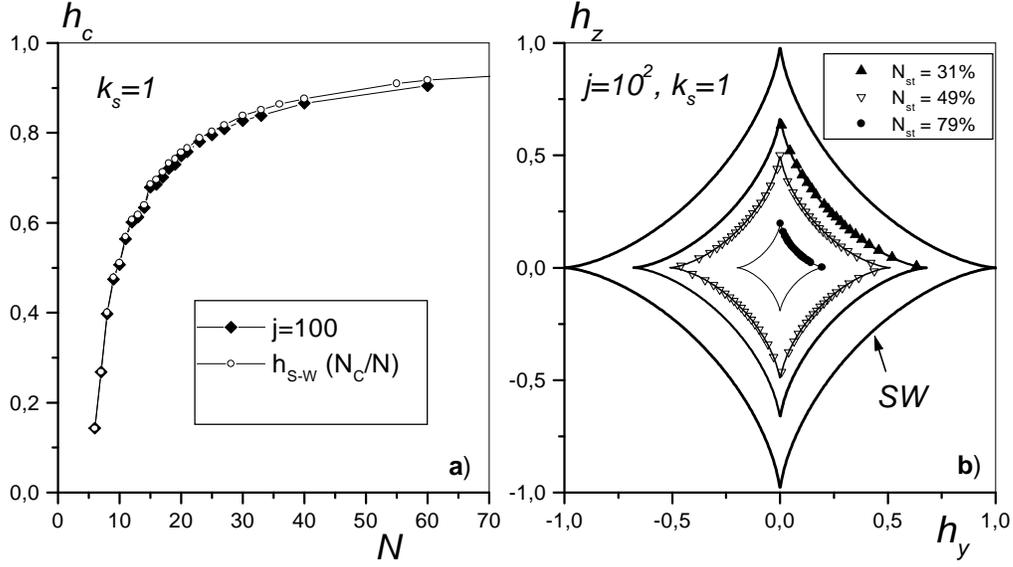} \vspace{-2cm}
\caption{a) (in diamonds) Switching field for $k_{s}=1,\,j=10^{2}$ versus
the particle's diameter $N$. (in circles) SW switching field multiplied by $%
N_{c}/\mathcal{N}$. b) Astroid for $k_{s}=1,\,j=10^{2}$ for different values
of the surface-to-volume ratio $N_{st}\equiv N_{s}/\mathcal{N}$. }
\label{fig2_jap}
\end{figure}

For small values of $k_{s}/j\sim 0.01$ the TSA model renders hysteresis
loops and limit-of-metastability curves that scale with the SW results for
all values of the angle $\psi $ between the core easy axis and the applied
field, the scaling constant being $\mathcal{N}_{c}/\mathcal{N}<1$, see in
Fig.~\ref{fig2_jap}. %
%

\begin{figure}[tbp] \includegraphics[angle=-90,width=15cm]{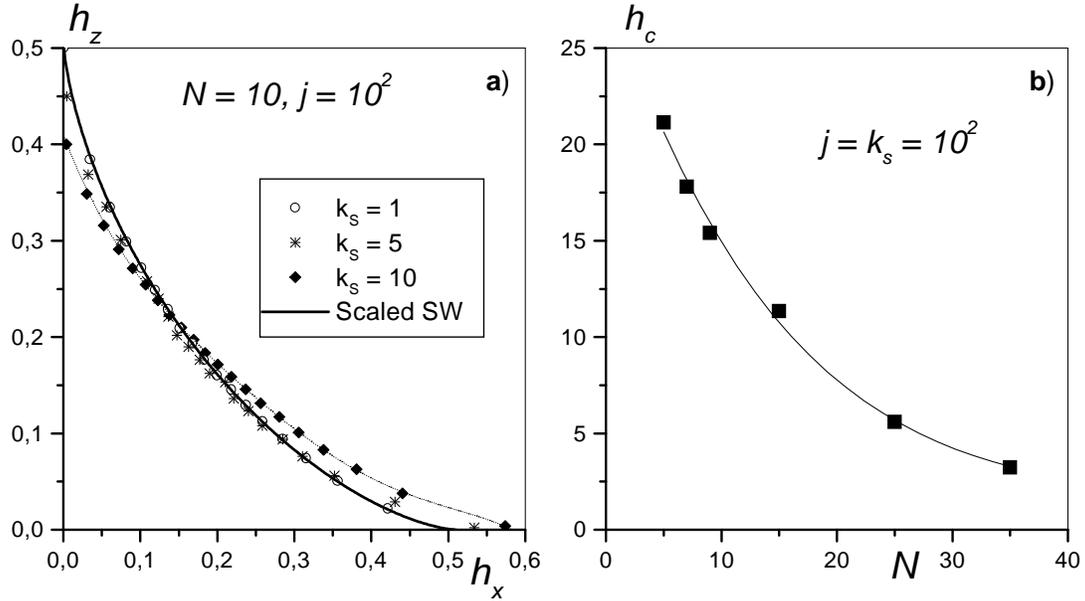}
\vspace{-2cm} \caption{a) Astroid for $j=10^{2}$, $\mathcal{N}=360$ and
different values of surface anisotropy constant $k_{s}$. The full dark line is
the SW astroid scaled with $N_{c}/\mathcal{N}$, but the dotted line is only a
guide for the eye. b) Switching field versus the particle's diameter $N$ for
$\protect\psi=0,\,j=k_{s}=10^{2}$. }
\label{fig3_jap}
\end{figure}

\begin{figure}[tbp]
\centerline{\includegraphics[angle=-90, width=12cm]{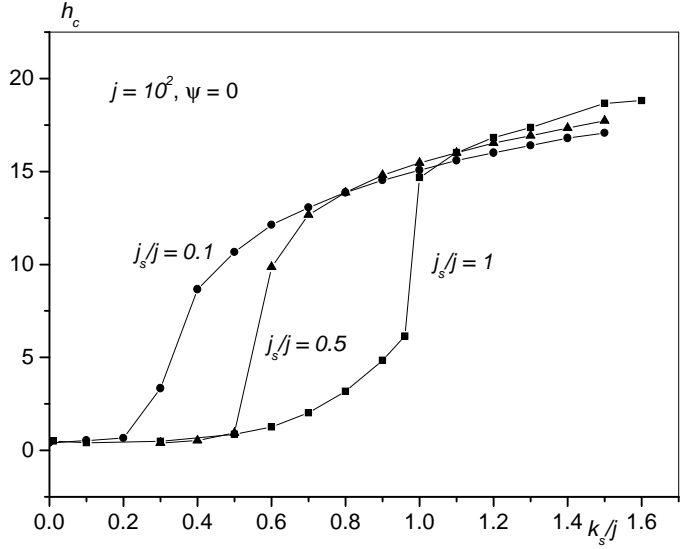}} \vspace{%
-0.75cm}
\caption{Critical field versus the surface anisotropy constant for $\protect
\psi =0$, and different values of surface-to-core ratio of exchange
couplings; $N=10$.}
\label{hcvsks2}
\end{figure}

For larger values of $k_{s}/j$, but $k_{s}/j\lesssim 0.2$, we still have the
same kind of scaling but the corresponding constant now depends on the
angle $\psi$ between the core easy axis and field direction. This is reflected
by a deformation of the limit-of-metastability curve (see Fig.~\ref{fig3_jap}a).
More precisely, the latter is depressed in the core easy direction and enhanced
in the perpendicular direction. However, there is still only one jump in the
hysteresis loop implying that the magnetization reversal can be considered as
uniform.
%
%
For $k_{s}/j\gtrsim 1$, there appear multiple steps in the hysteresis loop
associated with the switching of spin clusters. It makes the hysteresis loop
both qualitatively and quantitatively different from those of the SW model,
as the magnetization reversal can no longer be considered as uniform. In
addition, in the present case, there are two more new features: the values
of the switching field are much higher than in SW model, and more
importantly, its behavior as a function of the particle's size is opposite
to that of the previous cases (compare Figs.~\ref{fig2_jap}a and
\ref{fig3_jap}b).
More precisely, in this case one finds that
this field increases when the particle's size is lowered. This is in
agreement with the experimental observations in nanoparticles (see, e.g., \cite
{chesorklahad95prb} for cobalt particles).
The whole situation is summarized in Fig.~\ref{hcvsks2} where we plot the
critical field $h_{c}$ as a function of $\tilde{k}_{s}\equiv k_{s}/j$ for
different values of the surface-to-core ratio of the exchange coupling.
%
%
For large values of $k_{s},$ surface spins are aligned along their radial
easy axes, and because of strong exchange coupling they also drive core
spins in their switching process, which requires a very strong field to be
completed. The value of $k_{s}$ where $h_{c}$ jumps up (e.g., $=1$ for $%
J_{s}/J=1$) marks the passage from a regime where scaling with the SW
results is possible (either with a $\psi $-dependent or independent
coefficient) to the second regime where this scaling is no longer possible
because of completely different switching processes.

To estimate $K_{s}$ and the critical field, consider a 4 nm cobalt particle
of fcc crystal structure, for which the lattice spacing is $a=3.554\,\AA $,
and there are 4 cobalt atoms per unit cell. The (bulk) magneto-crystalline
anisotropy is $K_{c}\simeq 3\times 10^{-17}$ erg/spin or $2.7\times 10^{6}$
erg/cm$^{3}$, and the saturation magnetization is $M_{s}\simeq 1422$ emu/cm$%
^{3}$. The switching field is given by $H_{c}=(2K_{c}/M_{s})h_{c}$. For $%
\psi =0$, $\tilde{k}_{s}^{c}=1$ and $h_{c}=15$, so $H_{c}\simeq 6\,$T. On
the other hand, $\tilde{k}_{s}^{c}=1$ means that the effective exchange
field experienced by a spin on the surface is of the order of the anisotropy
field, i.e. $zSJ/2\sim 2K_{s}$. Then using $J\simeq 8\,$mev we get $%
K_{s}\simeq 5.22\times 10^{-14}\,$erg/spin, or using the area per surface
spin (approximately $a^{2}/8$), $K_{s}\simeq 5\,$erg/cm$^{2}$. For the case
of $\psi =\pi /4$, $\tilde{k}_{s}^{c}\simeq 0.2$ and $h_{c}\simeq 0.3$,
which leads to $H_{c}\simeq 0.1\,$T and $K_{s}\simeq 1.2\times 10^{-14}\,$%
erg/spin or $1.2\,$erg/cm$^{2}$.

\subsection{\label{sec:NSAKs}Surface contribution to the energy of magnetic
nanoparticles: NSA}

We calculate the contribution of the NSA \cite{nee54jpr} to the effective
anisotropy of magnetic nanoparticles of spherical shape cut out of a simple
cubic lattice. The effective anisotropy arises because of deviations of
atomic magnetic moments from collinearity and dependence of the energy on
the orientation of the global magnetization with respect to crystallographic
directions. We show \cite{garkac03prl} that the result is second order in
the NSA constant, scales with the particle's volume, and has cubic symmetry
with preferred directions $[\pm 1,\pm 1,\pm 1]$.

As shown many times before, as the size of the magnetic particle decreases,
surface effects become more and more pronounced. In many cases surface
atoms yield a contribution to the anisotropy energy that scales with
particle's surface, i.e., to the \emph{effective} volume anisotropy
decreasing with the particle's linear size $R$ as $K_{V,\mathrm{eff}%
}=K_{V}+K_{S}/R,$ as was observed in a number of experiments (see, e.g.,
Refs.~\cite{resetal98prb}, \cite{chekitokashi99jap}).
%
\begin{figure}[t]
\centerline{\includegraphics[angle=-90,width=8cm]{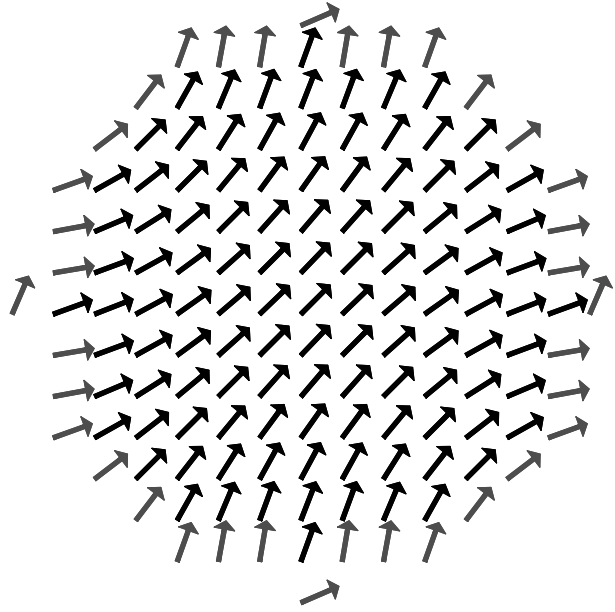}}
\caption{Magnetic structure in the plane $z=0$ of a spherical nanoparticle
of linear size $N=15$ with $L/J=2$ for the global magnetization directed
along [110].}
\label{nsa-structure}
\end{figure}
%
The $1/R$ surface contribution to $K_{V,\mathrm{eff}}$ is in accord with the
picture of all magnetic atoms tightly bound by the exchange interaction
while only surface atoms feel the surface anisotropy. This is definitely
true for magnetic films where a huge surface contribution to the effective
anisotropy has been observed. The same holds for cobalt nanoclusters of the
form of truncated octahedrons \cite{jametal01prl} where contributions from
different faces, edges, and apexes compete, resulting in a nonzero, although
significantly reduced, surface contribution to $K_{V,\mathrm{eff}}$.

However, for symmetric particle's shapes such as cubes or spheres, the
symmetry leads to vanishing of this first-order contribution. In this case
one has to take into account deviations from the collinearity of atomic
spins that result from the competition of the surface anisotropy and
exchange interaction $J$. The resulting magnetic structures (for the
simplified radial SA model) can be found in \cite{dimwys94prb}, \cite
{labetal02jap}, \cite{kacdim02prbjap} (see also Fig.~\ref{nsa-structure} for
the NSA). In the case $L\gtrsim J$ deviations from collinearity are very
strong, and it is difficult, if not impossible, to characterize the particle
by a global magnetization suitable for the definition of the effective
anisotropy. For $L\ll J$ the magnetic structure is nearly collinear with
small deviations that can be computed perturbatively in $L/J\ll 1$. The
global magnetization vector $\mathbf{m}_{0}$ [this is the same as the vector
defined in (\ref{mdef})] can be used to define the anisotropic energy of the
whole particle. The key point is that deviations from collinearity, and
thereby the particle's energy, depend on the orientation of $\mathbf{m}_{0}$%
, even for a particle of a spherical shape, due to the crystal lattice. To
illustrate this idea, we neglect the bulk anisotropy and the DDI in the
Hamiltonian (\ref{hamgeneral}). For a sc lattice Eq.~(\ref{NSA}) reduces to
\begin{equation}
\mathcal{H}_{\mathrm{an}}^{(\mathrm{NSA})}=\sum_{i}\mathcal{H}_{\mathrm{an}%
,i}^{(\mathrm{NSA})},\qquad \mathcal{H}_{\mathrm{an},i}^{(\mathrm{NSA})}=%
\frac{L}{2}\sum_{\alpha =x,y,z}z_{i\alpha }s_{i\alpha }^{2},  \label{NSAsc}
\end{equation}
where $z_{i\alpha }=0,1,2$ are the numbers of available nearest neighbors of
the atom $i$ along the axis $\alpha .$ One can see that the NSA is in
general biaxial. For $L>0$ and $z_{i\alpha }=0<z_{i\beta }=1<z_{i\gamma }=2$
the $\alpha $-axis is the easy axis and the $\gamma $-axis is the hard axis.
If the local magnetic moments $\mathbf{s}_{i}$ are all directed along one of
the crystallographic axes $\alpha $, then the anisotropy fields $\mathbf{H}%
_{Ai}=-\partial \mathcal{H}_{Ai}/\partial \mathbf{s}_{i}$ are also directed
along $\alpha $ and are thus collinear with $\mathbf{s}_{i}$. Hence, at
least for $L\ll J$, there are no deviations from collinearity if the global
magnetization $\mathbf{m}_{0}$ is directed along one of the crystallographic
axes. For other orientations of $\mathbf{m}_{0}$, the vectors $\mathbf{s}%
_{i} $ and $\mathbf{H}_{Ai}$ are not collinear, since then at least two
components of $\mathbf{s}_{i}$ are non zero, and the transverse component of
$\mathbf{H}_{Ai}$ with respect to $\mathbf{s}_{i}$ causes a slight canting
of $\mathbf{s}_{i}$ and thereby a deviation from the collinearity of
magnetizations on different sites. This adjustment of the magnetization to
the surface anisotropy lowers the energy. As we shall see, this effect is
strongest for the [$\pm $1,$\pm $1,$\pm $1] orientations of $\mathbf{m}_{0}$%
. For both signs of $L$ these are easy orientations, whereas [$\pm 1,0,0],$ [%
$0,\pm 1,0],$ and [$0,0,\pm 1]$ are hard orientations.

%
\begin{figure}[t]
\includegraphics[angle=-90,width=7cm]{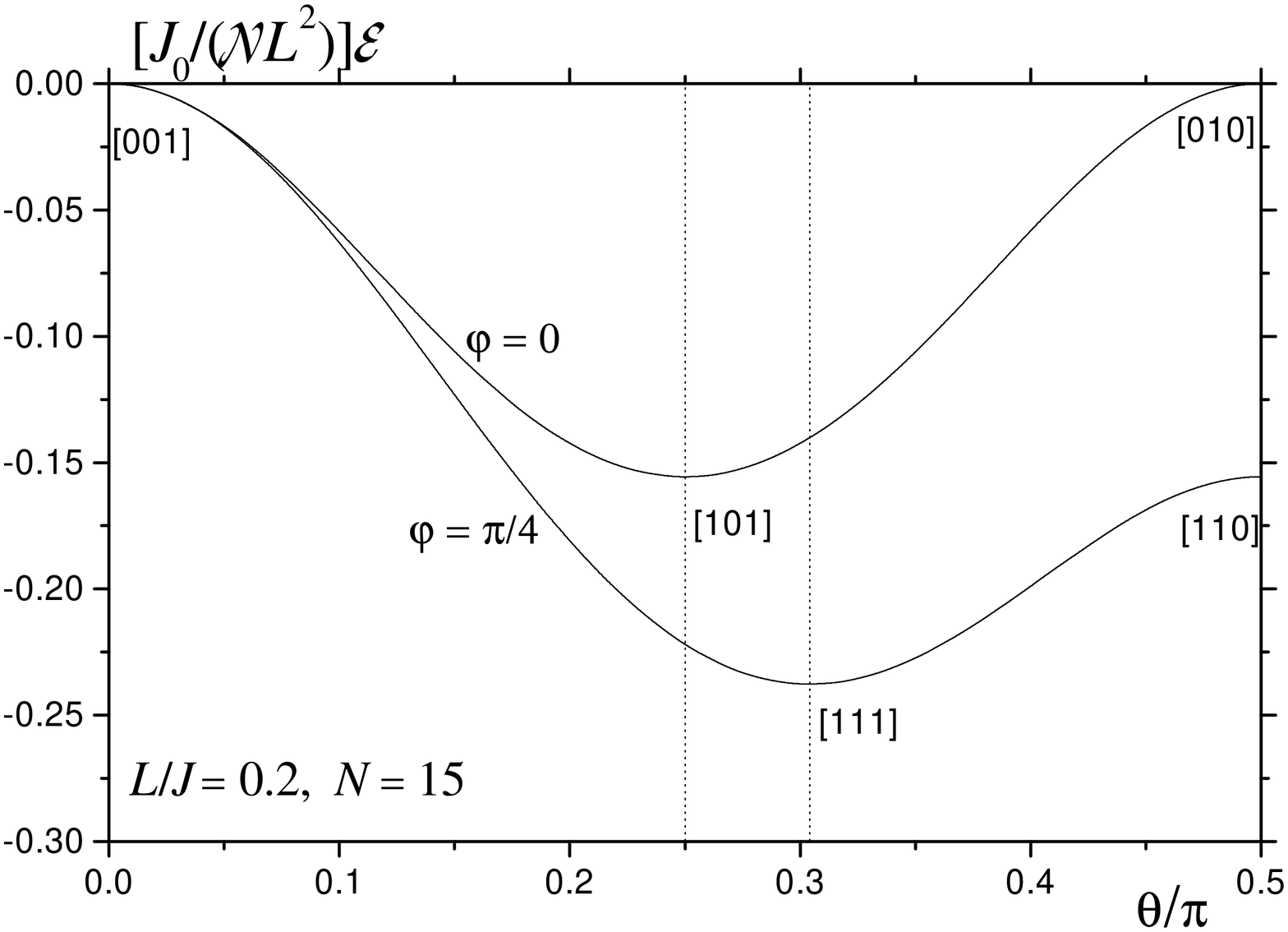} %
\includegraphics[angle=-90,width=7cm]{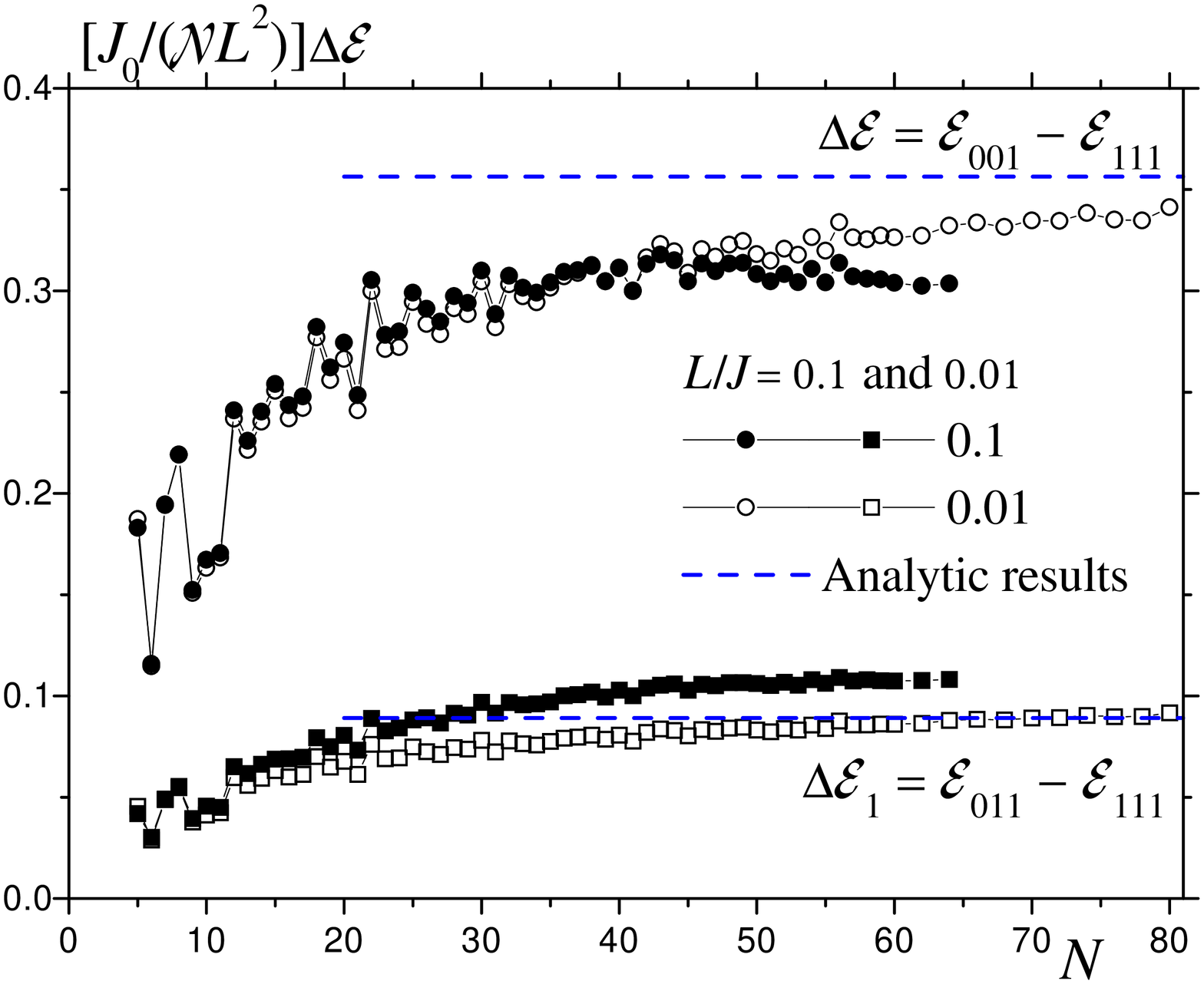}
\caption{Left: Effective anisotropy energy of the particle for different
orientations of its global magnetization showing cubic symmetry of Eq.~(\ref
{Ekappa}). Right: Differences of particle's energies between main
orientations of the global magnetization vs the particle size in the scaled
form for $L/J=0.1$ and 0.01. The scaling is valid for $N\lesssim J/L$, and
its violation for $L/J=0.1$ is seen in the right part of the figure. }
\label{nsa-etheta}
\end{figure}
%
To solve the problem numerically, we have to fix the global magnetization of
the particle in a desired direction $\mathbf{\nu }_{0}$ ($|\mathbf{\nu }%
_{0}|=1$) by using the energy function with a vector Lagrange multiplier $%
\mathbf{\lambda }$:
\begin{equation}
\mathcal{F}=\mathcal{H}-\mathcal{N}\mathbf{\lambda }{\cdot }\left( \mathbf{%
\nu }-\mathbf{\nu }_{0}\right) ,\qquad \mathbf{\nu }{\equiv }\frac{\sum_{i}%
\mathbf{s}_{i}}{\left| \sum_{i}\mathbf{s}_{i}\right| }.  \label{FFuncDef}
\end{equation}
To minimize $\mathcal{F}$ we numerically integrate the evolution equations
\begin{eqnarray}
\mathbf{\dot{s}}_{i} &=&-\left[ \mathbf{s}_{i}\times \left[ \mathbf{s}%
_{i}\times \mathbf{F}_{i}\right] \right] ,\qquad \mathbf{F}_{i}\equiv
-\partial \mathcal{F}/\partial \mathbf{s}_{i}  \nonumber  \label{LLEqs} \\
\mathbf{\dot{\lambda}} &=&\mathbf{\partial }\mathcal{F}/\partial \mathbf{%
\lambda }{=-}\mathcal{N}\left( \mathbf{\nu }-\mathbf{\nu }_{0}\right) .
\end{eqnarray}
starting from $\mathbf{s}_{i}=\mathbf{\nu }_{0}=\mathbf{m}_{0}$ and $\mathbf{%
\lambda }{=0,}$ until the stationary state is attained and an energy minimum
is found. Our numerical results for the magnetic energy of spherical
particles as a function of the orientation of the global magnetization are
shown in Fig.~\ref{nsa-etheta} (left). Fig.~\ref{nsa-etheta} (right) shows
differences between the basic directions [001], [011], and [111]. It is seen
that $\Delta E/\mathcal{N\rightarrow }\mathrm{const}$ for $N\rightarrow
\infty $ limit, i.e., $\Delta E$ scale with the particle's volume $V\propto
\mathcal{N}$.

To analytically solve the problem in the continuous limit, we replace in
Eq.~(\ref{NSAsc}) the number of nearest neighbors of a surface atom by its
average value
\begin{equation}
z_{i\alpha }\Rightarrow \overline{z}_{i\alpha }=2-|n_{\alpha }|/\max \left\{
|n_{x}|,|n_{y}|,|n_{z}|\right\} .  \label{zAvr}
\end{equation}
Here $n_{\alpha }$ is the $\alpha $-component of the normal to the surface $%
\mathbf{n}$. The surface-energy density can then be obtained by dropping the
constant term and multiplying Eq.~(\ref{NSAsc}) by the surface atomic
density $f(\mathbf{n})=\max \left\{ |n_{x}|,|n_{y}|,|n_{z}|\right\} $:
\begin{equation}
E_{S}(\mathbf{m,n})=-\frac{L}{2}\left[ |n_{x}|m_{x}^{2}+|n_{y}|m_{y}^{2}+%
\left| n_{z}\right| m_{z}^{2}\right] .  \label{ESDef}
\end{equation}
At equilibrium the Landau-Lifshitz equation reads
\begin{equation}
\mathbf{m\times H}_{\mathrm{eff}}=0,\qquad \mathbf{H}_{\mathrm{eff}}=\mathbf{%
H}_{A}+J\Delta \mathbf{m.}  \label{LLEqEqui}
\end{equation}
For small deviations from collinearity one can seek for its solution in the
form
\begin{equation}
\mathbf{m(r)\cong m}_{0}+\mathbf{\psi }(\mathbf{r,m}_{0}),\qquad \psi \equiv
|\mathbf{\psi }|\ll 1  \label{mAnsatz}
\end{equation}
where $\mathbf{\psi }$ is the solution of the internal Neumann boundary
problem
\begin{eqnarray}
&&\Delta \mathbf{\psi }=0,\qquad \left. \frac{\partial \mathbf{\psi }}{%
\partial r}\right| _{r=R}=\mathbf{f(m,n)}  \nonumber \\
&&\mathbf{f}=-\frac{1}{J}\left[ \frac{dE_{S}(\mathbf{m,n})}{d\mathbf{m}}%
-\left( \frac{dE_{S}(\mathbf{m,n})}{d\mathbf{m}}\cdot \mathbf{m}\right)
\mathbf{m}\right] .  \label{psiNeumann}
\end{eqnarray}
This equation can be solved with the help of the Green function $G(\mathbf{%
r,r}^{\prime })$ (see Ref. \cite{garkac03prl}), and the final result for the
second-order energy contribution is
\begin{equation}
\mathcal{E}_{2}\cong \frac{1}{2\pi J}\int \int_{S}d^{2}\mathbf{r}d^{2}%
\mathbf{r}^{\prime }G(\mathbf{r,r}^{\prime })E_{S}(\mathbf{m,n})E_{S}(%
\mathbf{m,n}^{\prime }).  \label{EFinal}
\end{equation}
Taking into account the cubic symmetry and computing numerically a double
surface integral one can write the result of Eq.~(\ref{EFinal})\ as
\begin{equation}
\mathcal{E}_{2}\cong \kappa \frac{L^{2}\mathcal{N}}{J_{0}}\left(
m_{x}^{4}+m_{y}^{4}+m_{z}^{4}\right) ,\qquad \kappa =0.53465,  \label{Ekappa}
\end{equation}
where $J_{0}=zJ=6J.$ This defines the large-$N$ asymptotes in Fig.~\ref
{nsa-etheta} (right) shown by the horizontal lines.

The analytical results above is valid for particle sizes $N$ in the range
\begin{equation}
1\ll N\ll J/L.  \label{NApplReg}
\end{equation}
The lower boundary is the applicability condition of the continuous
approximation. Since the surface of a nanoparticle is made of atomic
terraces separated by atomic steps, each terrace and each step with its own
form of NSA [see Eq.~(\ref{NSAsc})], the variation of the local NSA along
the surface is very strong. Approximating this variation by a continuous
function according to Eq.~(\ref{zAvr}) requires pretty large particle sizes $%
N$. This is manifested by a slow convergence to the large-$N$ results in
Fig.~\ref{nsa-etheta} (right). The upper boundary in Eq.~(\ref{NApplReg}) is
the applicability condition of the linear approximation in $\mathbf{\psi }$.
For $N\gtrsim J/L$ deviations from the collinear state are strong, and the
effective anisotropy of a magnetic nanoparticle cannot be introduced.

As we have seen in Eq.~(\ref{Ekappa}), the contribution of the surface
anisotropy to the overall anisotropy of a magnetic particle scales with its
volume $V\propto N^{3}\sim \mathcal{N}$. This surprising result is due to
the penetration of perturbations from the surface deeply into the bulk. If a
uniaxial bulk anisotropy $K_{c}$ was present in the system, perturbations
from the surface would be screened at the bulk correlation length (or the
domain-wall width) $\delta \sim \sqrt{J/K_{c}}$. Then for $N\gtrsim \delta $
the contribution of the surface anisotropy to the overall anisotropy would
scale as the surface: $\mathcal{E}_{2}\sim \left( L^{2}/J\right) N^{2}\delta
$. For not too large particles, $N\lesssim \delta $, contributions of both
anisotropies to the anisotropic energy are additive and scale as the volume.
If the bulk anisotropy is cubic, both contributions have the same cubic
symmetry [c.f. Eqs.~(\ref{ca_xyz}) and (\ref{Ekappa})], and the experiment
should provide a value of the effective cubic anisotropy different form the
bulk value \cite{jametal01prl}. For the uniaxial bulk anisotropy, the two
contributions have different functional forms. Even if the bulk anisotropy
is dominant so that the energy minima are realized for $\mathbf{m}\Vert
\mathbf{e}_{z}$, the surface anisotropy makes the energy dependent on the
azimuthal angle $\varphi $. This modifies particle's energy barrier by
creating saddle points and strongly influences the process of thermal
activation \cite{garkencrocof99pre}.

For small deviations from the cubic or spherical shape, i.e., for weakly
elliptic or weakly rectangular particles, there should emerge a
corresponding weak first-order contribution $\mathcal{E}_{1}$ that would add
up with our second-order contribution. For an ellipsoid with axes $a$ and $%
b=a(1+\epsilon ),\epsilon \ll 1$, the anisotropy energy scales with the
surface, $\mathcal{E}_{1}\sim L\mathcal{N}^{2/3}\epsilon m_{z}^{2}$ [cf.
Eq.~(\ref{Ekappa})], so that
\begin{equation}
\frac{\mathcal{E}_{2}}{\mathcal{E}_{1}}\sim \frac{L}{J}\frac{N}{\epsilon }
\label{E2E1Ratio}
\end{equation}
can be large even for $L/J\ll 1.$

The N\'{e}el constant $L$ is in most cases poorly known. However, for
metallic Co \cite{chubaloha94prb} quotes the value of surface anisotropy $%
-1.5\times 10^{8}$ erg/cm$^{3}$, i.e., $L\sim -10$ K. This is much smaller
than $J\sim 10^{3}$ K, which makes our theory valid for particle sizes up to
$N\sim J/L\sim 100$, according to Eq.~(\ref{NApplReg}). For this limiting
size one has $\mathcal{E}_{2}/\mathcal{E}_{1}\sim 1/\epsilon $ that is large
for nearly spherical particles, $\epsilon \ll 1$.

\subsection{\label{sec:TE+SE}Surface effects on the magnetization of a
nanoparticle at $T>0$: MC}

In this section, we consider more realistic model of round-shaped (spherical
or ellipsoidal) nanoparticle of simple cubic or spinel crystalline
structure, uniaxial or cubic anisotropy in the core, and transverse or
N\'{e}el anisotropy on the surface, described by the Hamiltonian defined by
Eqs.~(\ref{hamgeneral}), (\ref{uaa}), (\ref{ca_xyz}), and (\ref{NSA}). Using
various techniques explained above, we compute the magnetization,
induced and intrinsic, as a function of temperature and applied magnetic
field, for different values of the surface anisotropy constant and exchange
coupling~\cite{kacetal00epjb}\cite{kacetal00jmmm}. We shall mainly focus on
novel features stemming from the combination of anisotropy, field and
temperature effects on the magnetization.

%
\begin{figure}[h]
\includegraphics[angle=-90, width=7.5cm]{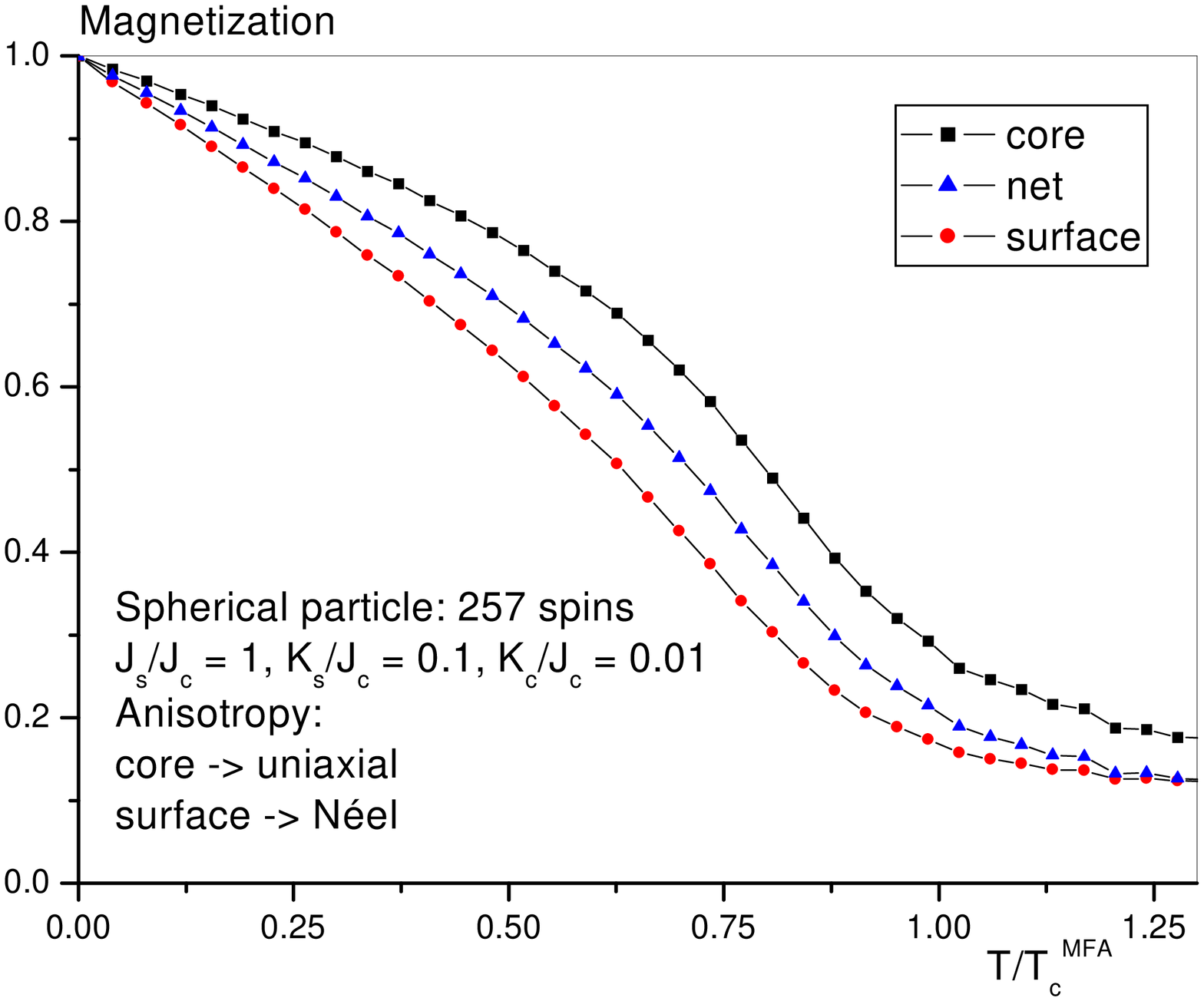} %
\includegraphics[angle=-90, width=7.5cm]{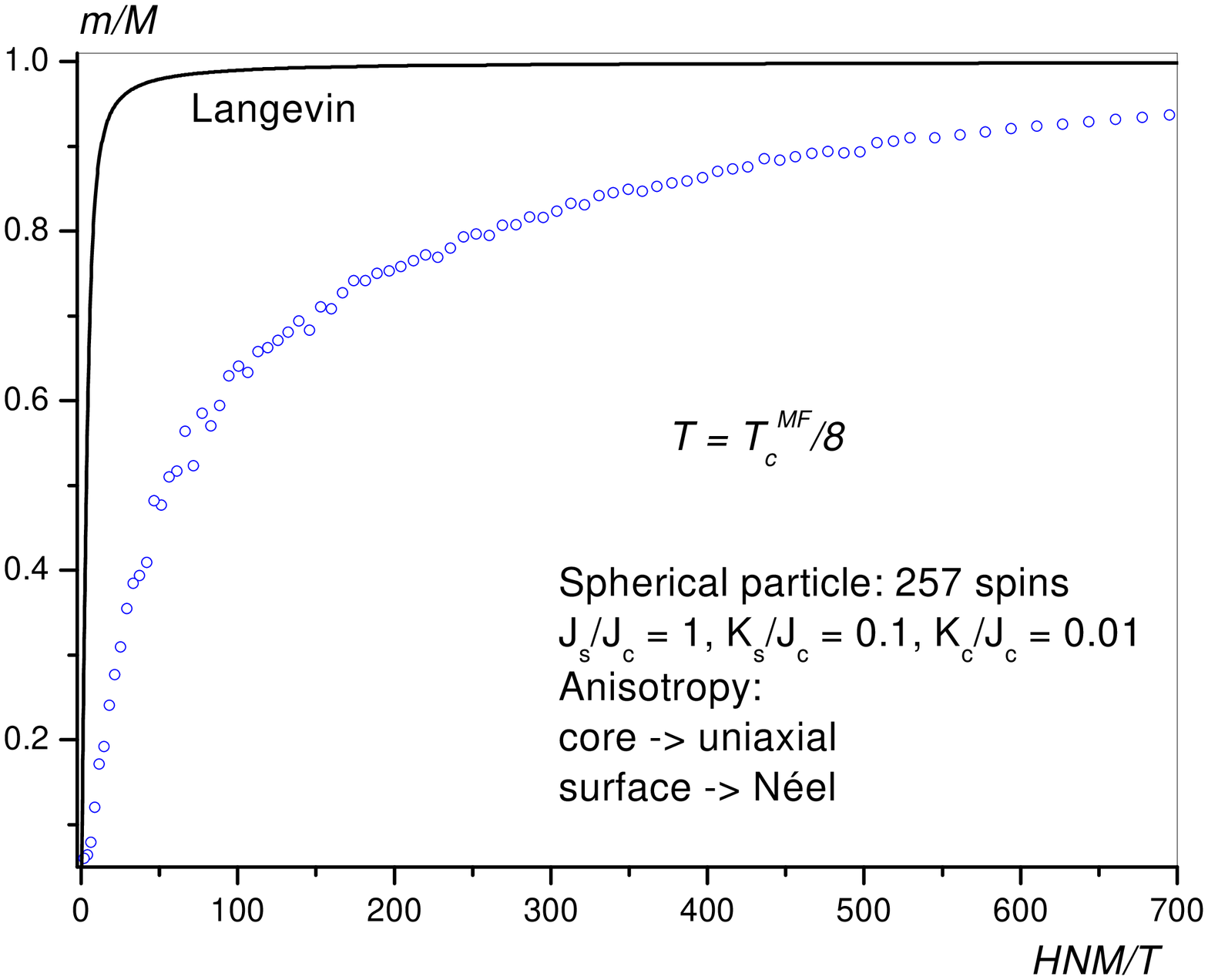}
\caption{Left: Core, surface and net magnetizations as functions of
temperature. Right: Circles: Scaled graph for the induced magnetization $m$
as a function of $HNM/T$, where $M$ is the intrinsic magnetization. Full
line: Langevin function $\mathit{L}(x)=\coth x-1/x$. }
\label{magT_N9_neel}
\end{figure}
%

\subsubsection{\label{sec:fmp_nsa}Ferromagnetic particles with N\'{e}el's
surface anisotropy}

In our simulations we consider the core ferromagnetic coupling $J_{c}$ and
uniaxial anisotropy $K_{c}=0.01J_{c}.$ On the surface we adopt $J_{s}=J_{c}$%
, while the anisotropy is given by the N\'{e}el expression (\ref{NSA}) with
constant $K_{s}=0.1J_{c}$. We ignore the DDI for simplicity. In particular,
we are interested in how anisotropy affects the superparamagnetic relation (%
\ref{spmrelation}) that has been shown to hold at all temperatures below $%
T_{c}$ for isotropic systems. As no analytical calculations are possible
here, we resort to the Monte Carlo technique with global spin rotations. In
Fig.~\ref{magT_N9_neel} (left) we plot the core, surface and net
magnetizations of a nanoparticle of $257$ spins, as functions of temperature
in zero magnetic field. These results do confirm what was obtained from the
spherical model [see Fig.~\ref{sef_t3d} (left)] for isotropic box-shaped
systems, namely that boundary effects suppress the magnetization, and here
we see that this effect is enhanced by the SA.A drastic effect of the SA is
clearly seen in the field dependence of the scaled induced magnetization $m$
as shown in Fig.~\ref{magT_N9_neel} (right) at $T=T_{c}^{MFA}/8$. In the
presence of a strong SA the superparamagnetic relation (\ref{spmrelation})
is no longer valid, and $m/M$ strongly deviates from the Langevin function.

%
\begin{figure}[h]
\includegraphics[width = 5.5cm, angle=-90]{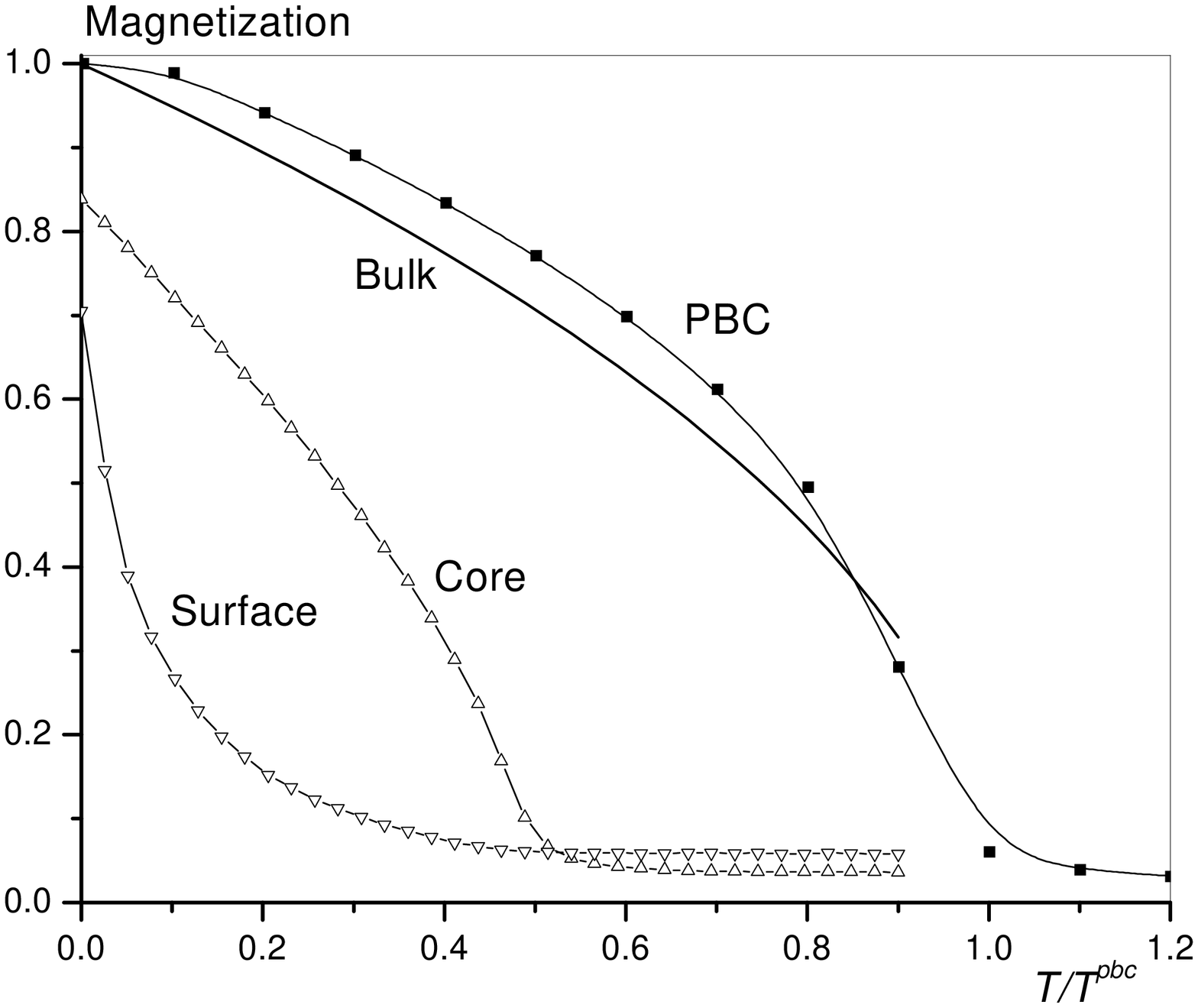} %
\includegraphics[width = 5.5cm, angle=-90]{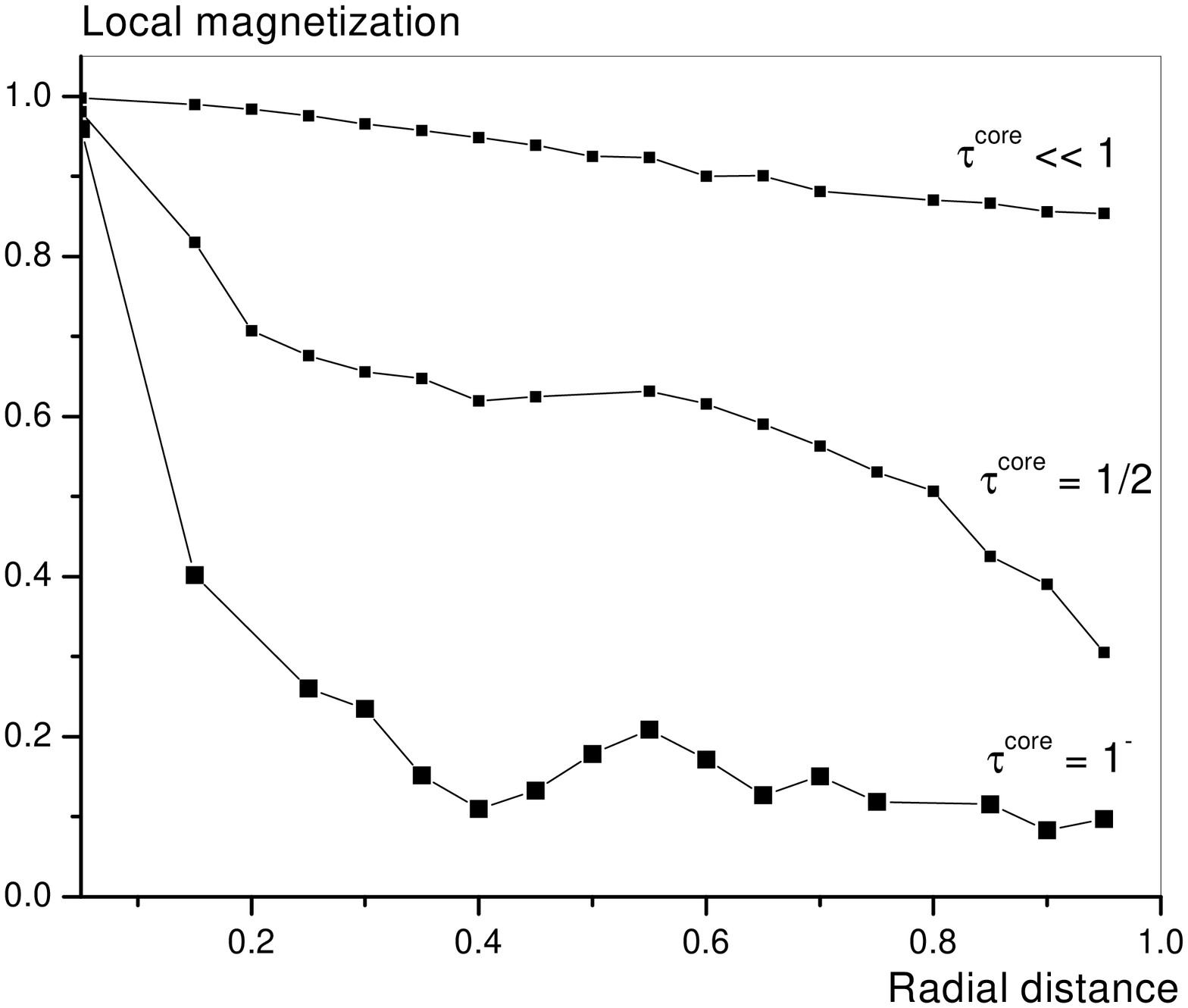}
\caption{Left: Temperature dependence of the surface and core magnetizations
for $\mathcal{N}=3766$, magnetization of the bulk system, and that of the
cube with the spinel structure and periodic boundary conditions (PBC) with $%
\mathcal{N}=40^{3}$. Right: Spatial variation of the net magnetization of a
spherical nanoparticle of $3140$ spins, as a function of the normalized
particle radius, for $\protect\tau ^{core}\equiv T/T_c^{\mathrm{core}}\ll 1$,
and $\protect\tau^{core}=0.5$, $\protect\tau ^{core}\simeq 1^{-}$. }
\label{Barce3}
\end{figure}
%

\subsubsection{\label{sec:GAMMA}Maghemite ($\protect\gamma $-Fe$_{2}$O$_{3}$%
) nanoparticles \ }

In this section we deal with the ferrimagnetic maghemite nanoparticles ($%
\gamma $-Fe$_{2}$O$_{3}$) having spinel crystalline structure, summarizing
the results of \cite{kacetal00epjb}, \cite{kacetal00jmmm}. This time we
include the DDI in the Hamiltonian of Eq.~(\ref{hamgeneral}). The bulk
anisotropy in such materials is cubic, still we are using a uniaxial
anisotropy to simplify the study of effects that are of more interest to us
here. We consider maghemite particles of various sizes ($\mathcal{N}\simeq
10^{3}-10^{5}$ that correspond to a radius of 2-3.5 nm) and with the
physical properties in the core (spinel crystal structure with lacuna,
exchange and dipolar interactions, anisotropy constant, etc.) as those of
the bulk, except that the anisotropy is taken as uniaxial. We use the TSA
model with $K_{s}=0.06$ erg/cm$^{2}$. All spins in the core and on the
surface are identical but interact via different couplings depending on
their locus in the lattice. We assume that the exchange interactions between
the core and surface spins are the same as those inside the core. Although
we treat only the crystallographically ``ideal'' surface, we do allow for a
scatter in the exchange constants on the surface. In contrast, in Refs. \cite
{kodber99prb} it was assumed that all exchange interactions are
the same but there was postulated the existence of a fraction of missing
bonds on the surface.

In Fig.~\ref{Barce3} (left), we see that the surface magnetization decreases
more rapidly than the core magnetization as the temperature increases.
Moreover, it is seen that even the (normalized) core magnetization per site
does not reach its saturation value of $1$ at very low temperatures, which
shows that the magnetic order in the core is disturbed by the surface. This
may also be due to lacuna in the spinel structure. In Fig.~\ref{Barce3}
(right) we plot the spatial evolution of the local magnetization from the
center to the border of the particle, at different temperatures. At all
temperatures it decreases with distance from the center. At high
temperatures, the local magnetization exhibits a temperature-dependent jump,
and then continues to decrease. This indicates that there is a radius within
which the magnetization assumes relatively large values. This result agrees
with that of \cite{wil74} where this radius was called the \textit{magnetic
radius}.

\section{Conclusion}

We have demonstrated by different analytical and numerical methods the
importance of accounting for the magnetization inhomogeneities in magnetic
nanoparticles, especially in the presence of SA.  The
latter makes the magnetization inhomogeneous even at $T=0$ and in general
modifies the relation between the intrinsic and induced magnetizations. It also
changes the magnetization switching mechanism, since for large SA the
particle's spins switch cluster-wise.
For small SA we were able to calculate the
spin canting in the particle analytically and to obtain a novel second order
contribution to the particle's overall anisotropy. It remains to generalize this
result for nonzero bulk anisotropy. Another important task is to study dynamical
implications of the many-body effects in magnetic nanoparticles.

\end{document}